\documentclass[aps,prb,twocolumn,groupedaddress,superscriptaddress,showpacs,floatfix]{revtex4-1}
\usepackage{graphicx}

\usepackage{xcolor}
\usepackage{amsfonts}
\usepackage{amsmath}
\usepackage{amssymb}
\usepackage{rotating}
\usepackage{bm}

\usepackage[unicode=true, bookmarks=false,breaklinks=false,pdfborder={0 0 1},backref=false,colorlinks=false]{hyperref}
\hypersetup{hypertex}   % gera link para referencias ao longo do texto
\usepackage{breakurl}
\hypersetup{backref,colorlinks=true,citecolor=blue,linkcolor=blue,filecolor=blue,urlcolor=blue}
\usepackage[english]{babel}

\begin{document}

\title{A tight-binding model for the band dispersion in rhombohedral topological insulators over the whole Brilluoin zone}

\author{Carlos Mera Acosta}
\email[*]{acosta@if.usp.br}
\affiliation{Instituto de F\'isica, Universidade de S\~ao Paulo, CP 66318, 05315-970, S\~ao Paulo, SP, Brazil}
\affiliation{Brazilian Nanotechnology National Laboratory, CP 6192, 13083-970, Campinas, SP, Brazil}

\author{Matheus P. Lima} 
\affiliation{Departamento de F\'isica, Universidade Federal de S\~ao Carlos, CP 676, 13565-905, S\~ao Carlos, SP, Brazil}

\author{Ant\^onio J. R. da Silva} 
\affiliation{Instituto de F\'isica, Universidade de S\~ao Paulo, CP 66318, 05315-970, S\~ao Paulo, SP, Brazil}
\affiliation{Laborat\'orio Nacional de Luz S\'{\i}ncrotron, CP 6192, 13083-970, Campinas, SP, Brazil}

\author{A. Fazzio} 
\affiliation{Instituto de F\'isica, Universidade de S\~ao Paulo, CP 66318, 05315-970, S\~ao Paulo, SP, Brazil}
%\affiliation{Centro de Ci\^encias Naturais e Humanas, Universidade Federal do ABC, Santo Andr\'e, 09210-170 S\~ao Paulo, Brazil}
\affiliation{Brazilian Nanotechnology National Laboratory, CP 6192, 13083-970, Campinas, SP, Brazil}

\author{C. H. Lewenkopf} 
\affiliation{Instituto de F\'isica, Universidade Federal Fluminense, 24210-346 Niter\'oi, Brazil}

%\date{\today}

\begin{abstract}
We put forward a tight-binding model for rhombohedral topological insulators materials with the
space group $D^{5}_{3d}(R\bar{3}m)$. The model describes the bulk band structure of these materials 
over the whole Brillouin zone. Within this framework, we also describe the topological nature of 
surface states, characterized by a Dirac cone-like dispersion and the emergence of surface projected bulk states 
near to the Dirac-point in energy. We find that the breaking of the $R_{3}$ symmetry as one moves away from 
the $\Gamma$ point has an important role in the hybridization of the $p_x$, $p_y$, and $p_z$ atomic orbitals. 
In our tight-binding model, 
the latter leads to a band mixing matrix element ruled by a single parameter.
We show that our model gives a good description of the strategies/mechanisms proposed in the literature to
eliminate and/or energy shift the bulk states away from the Dirac point, such as stacking faults and the introduction 
of an external applied electric field.
\end{abstract}


\maketitle

%%%%%%%%%%%%%%%%%%%%%%%%%%%%%%%%%%%%%%%%%%%%
\section{Introduction}
\label{sec:introduction}
%%%%%%%%%%%%%%%%%%%%%%%%%%%%%%%%%%%%%%%%%%%%

Topological insulator (TI) materials have attracted a lot of attention over the recent years 
\cite{annurev-conmatphys-062910-140432,RevModPhys.83.1057,RevModPhys.82.3045}. 
Their unusual metallic surface electronic structure on an inverted bulk band gap and the 
time reversal (TR) topological protection of these states, which forbids the backscattering, make 
TIs very fascinating materials \cite{annurev-conmatphys-062910-140432,RevModPhys.83.1057, 
RevModPhys.82.3045,PhysRevLett.105.166803,Zhang2009,PhysRevB.94.041302}.
Due to the advances in synthesis techniques\cite{Eremeev2012} and their simple mathematical
\cite{PhysRevB.82.045122} and computational modeling\cite{Yang2012}, Bi$_2$Se$_3$-like materials 
have been referred as the ``hydrogen atom'' of the 3DTI\cite{Xia2009}. These systems have been proposed
as platforms for spintronic devices based on the control of induced magnetic moment direction
\cite{PhysRevLett.109.076801}, surface barriers\cite{PhysRevB.90.205431}, and single-atom 
magnetoresistance\cite{Awadhesh2015}. 

In addition to the metallic surface topological protected states in a insulating bulk, experiments find that
Bi$_2$Se$_3$-like materials exhibit electronic scattering channels, attributed to the presence of 
bulk states near in energy to the Dirac point\cite{annurev-conmatphys-062910-140432,Zhang2009, 
PhysRevLett.107.056803}. 
These ubiquitous bulk states are believed to prevent the observation of the expected unusual 
electronic and transport properties governed by surface states in 3DTIs\cite{PhysRevLett.107.056803, 
Brahlek201554,deVries2017}.

First principles $GW$ calculations for surface states~\cite{Louie2012,PhysRevB.92.201404,PhysRevB.93.205442} 
show that bulk states of 
Bi$_{2}$Se$_{3}$ thin films are shifted below the Dirac point, while this is not the case for 
Bi$_{2}$Te$_{3}$. In contrast, other bulk band structure calculations show  that there is barely 
any energy separation between the Dirac point and the bulk valence band 
maximum~\cite{PhysRevB.93.205442,Nechaev2013a,Aguilera2013a}.
This is at odds with recent experimental results \cite{deVries2017} that, by investigating 
Shubnikov-de Haas oscillations in this material, showed the coexistence of surface states 
and bulk channels with high mobility.

In order to obtain insight on this problem and understand the experimentally observed 
magnetotransport properties of thin films of rhombohedral TI materials, one needs an effective 
model capable of describing both the topological surface states as well as the bulk ones over the 
whole Brillouin zone.
In addition, the effective Hamiltonian has to account for the presence of external magnetic fields 
and be amenable to model disorder effects, which is beyond the scope of first principle methods. 
The main purpose of this paper is to put forward a tight-binding model that fulfills these characteristics.

Based on symmetry  properties and $\boldsymbol{k}\cdot\boldsymbol{p}$ perturbation theory, 
Zhang and collaborators \cite{PhysRevB.82.045122} derived a Dirac-like Hamiltonian model 
describing the low energy band structure around the $\Gamma$-point of Bi$_2$Se$_3$-like 3DTIs. 
Subsequently\cite{PhysRevB.84.115413}, a tight-binding effective model has been proposed
to describe the Brillouin of these systems, realizing both strong and weak TIs. 
However, the basis set used in such works fails to account for bulk states in the energy vicinity of 
the Dirac point and, hence, their effect on the electronic properties.  

Here, 
%following the approach put forward in Ref.~\onlinecite{PhysRevB.84.115413},
we propose an effective tight-binding model that provides insight on the above mentioned bulk 
states close to the Fermi energy that potentially spoil the bulk-boundary duality. In the 
presence of disorder these states can mix with the surface ones, quenching the 
topological properties of the material. 
% Further, we find that breaking the in-plane full rotation symmetry down to threefold rotation symmetry 
% has an important role in the hybridization of the $p_{x}p_{y}$ and $p_z$ Se atomic orbitals. 
% This hybridization leads to a band repulsion 
%
% \CAIO{Choose other words: referee complaints about ``band repulsion".}
%
% that shifts the bulk states above the Fermi level. We find that the band repulsion 
%
% \CAIO{same here} 
%
% depends mainly on the magnitude of the matrix element that mixes the  $3/2$ and $1/2$ total 
% angular momentum effective states. These states are formed by combinations of the $p_{x}p_{y}$ 
% and $p_{z}p_{x}p_{y}$ atomic orbitals, respectively. 
% Thus, the substitution of Se atoms by S atoms lead to the elimination/energy shift of 
% the continuous bulk states.    
We also use our model to discuss some known mechanisms to cause an energy shift of the bulk states, 
such as, stacking faults~\cite{Seixas2013} and applying an external electric field ~\cite{PhysRevLett.105.266806}.

This paper is organized as follows. In Sec.~\ref{sec:tight-binding}, we derive a tight-binding model 
for Bi$_2$Se$_3$-type 3DTI materials, that is based on their crystal structure symmetries and 
reproduces the bulk {\it ab initio} band structure calculations, thus describing the continuous bulk states near 
the Fermi level. In Sec.~\ref{sec:thin_films} we calculate the surface modes and discuss the 
microscopic origin of the bulk states in these materials. In Sec.~\ref{sec:shift} we study mechanisms 
to eliminate and/or shift the bulk states below the Fermi surface. 
Finally, we present our conclusions in Sec. \ref{sec:conclusions}.
The paper also contains one Appendix containing a detailed technical description of the effective 
model and the tight-binding parameters for both Bi$_{2}$Se$_{3}$ andBi$_{2}$Te$_{3}$ compounds.

%%%%%%%%%%%%%%%%%%%%%%%%%%%%%%%%%%%%%%%%%%%
\section{TIGHT BINDING EFFECTIVE MODEL}
\label{sec:tight-binding}
%%%%%%%%%%%%%%%%%%%%%%%%%%%%%%%%%%%%%%%%%%%

%-------------------------------------   FIGURE 1  -----------------------------------------------------
\begin{figure*}
\includegraphics[width = 0.95\linewidth]{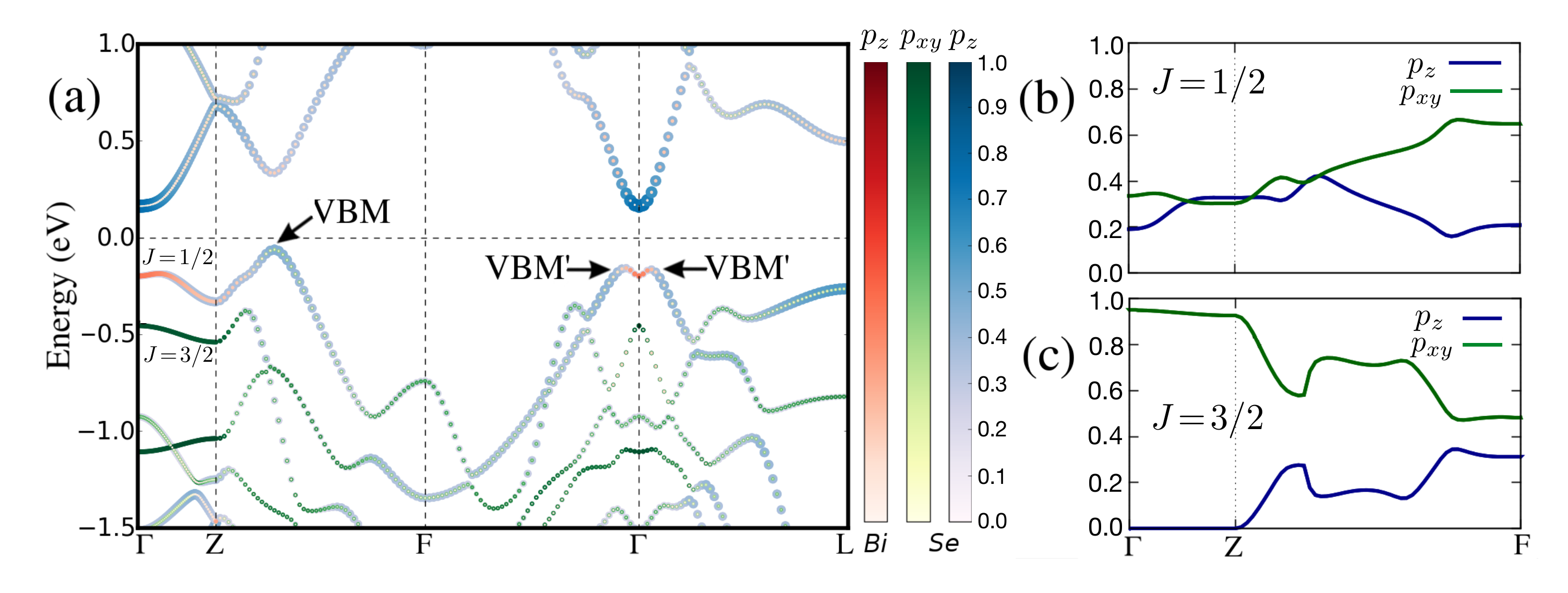}
\caption{(Color online) (a) Bulk band structure of Bi$_{2}$Se$_{3}$. The color code stands 
for the projections of the $p_{z}$ Bi orbitals (red), $p_{x}p_{y}$ Se orbitals (blue), and 
$p_{z}$ Se orbitals (green) in the wave function. The maximum and local maxima of the 
valence band are denoted by VBM and VBM', respectively. Panels (b) and (c) give the 
$p_{z}$ and $p_{x}p_{y}$ contributions of the $J=1/2$ and $J=3/2$ bands, respectively.}
\label{Int1}
\end{figure*}
%---------------------------------   END FIGURE 1 ---------------------------------------------------

We begin this section by reviewing the key symmetry arguments that allow one to 
obtain a simple effective tight-binding model for Bi$_{2}$Se$_{3}$-like 3DTIs. 
Next, we present the \textit{ab initio} electronic structure calculations on which 
our effective tight-binding model is based. 

The crystalline structure of Bi$_2$Se$_3$-like 3DTIs is formed by Quintuple-Layers (QL) 
characterized by $D_{3d}^{5}(R{\overline {3}}m)$ point group symmetries \cite{Zhang2009}.
The Bi$_2$Se$_3$ QL unit cell is composed by two bismuth and three selenium atoms 
\cite{Zhang2009}.
The QL-QL interaction is weak, mainly ruled by the Van der Waals-like interaction 
\cite{Zhang2009,PhysRevB.82.045122,Seixas2013}.
This allows one to model each QL unit cell by a triangular lattice site.
Following the approach presented in Ref.~\onlinecite{PhysRevB.82.045122}, the Bi$_2$Se$_3$ 
hexagonal unit cell is conveniently described by three triangular lattice layers stacked in the $z$ 
direction, instead of considering three QL unit cells.
This simple model preserves the symmetries of the $D_{3d}^{5}(R\overline{3}m)$ point group, 
namely: \textit{i}) threefold rotation symmetry $R_3$ along the $z$ axis,
 \textit{ii}) twofold rotation symmetry $R_2$ along the $x$ axis, 
 \textit{iii}) inversion symmetry $\mathcal{P}$, 
 and \textit{iv}) time-reversal symmetry $\mathcal{T}$.

It is well established \cite{PhysRevB.82.045122, Zhang2009} that the bulk wave 
function at the $\Gamma$ point can be accurately described by a set of few effective 
states $\{|\Lambda^{\tau}_{J},j_{z}\rangle\}$. 
Here, $\tau$ is the state parity, $J$ is the total angular 
momentum with projection $j_{z}$ on the $z$ axes, and $\Lambda$ labels the Bi and Se 
orbital contributions. 
We use these states to obtain an effective Hamiltonian that reproduces the bulk states of 
rhombohedral TIs calculated using {\it ab initio} methods. 

The first-principle calculations are performed within the Density Functional Theory (DFT) 
framework\cite{Capelle2006bird}, as implemented in the SIESTA 
code\cite{soler2002siesta}, considering the on-site approximation for the 
spin-orbit coupling\cite{fernandez2006site,PhysRevB.89.155438}. The Local Density 
Approximation (LDA)\cite{perdew1981self} is used for the exchange-correlation functional.

Figure~\ref{Int1} summarizes our \textit{ab initio} results for Bi$_{2}$Se$_{3}$.
The color code represents the contribution of the Bi and Se $p_{z}$ orbitals 
and the Se $p_{x}p_{y}$ atomic orbitals to the electronic structure.
The main orbital contributions are associated with $p$ orbitals corresponding to
$J=3/2, 1/2$ and $j_{z}=\pm3/2, \pm1/2$ states (Fig.~\ref{Int1}a).
To conserve the total angular momentum the $|\Lambda^{\pm}_{3/2},\pm 3/2\rangle$ effective 
states must be a linear combination of $p_x$ and $p_y$ orbitals, whereas the 
$|\Lambda^{\pm}_{J},\pm 1/2\rangle$ states correspond to a linear combination 
of all $p$ orbitals (Fig. \ref{Int1}b and Fig. \ref{Int1}c).
The symmetry properties of the $|\Lambda^{\tau}_{J},j_{z}\rangle$ states are discussed 
in Appendix A.

The bulk Valence Band Maximum (VBM) is located along the $Z\rightarrow F$ symmetry 
path, as shown in Fig.~\ref{Int1}a. 
In addition, one finds two local maxima, denoted by VBM', along the $F\rightarrow\Gamma$ and 
$\Gamma\rightarrow L$ 
% symmetry 
lines, both close to the $\Gamma$-point. 
In line with previous results\cite{PhysRevB.84.115413}, we observe that both VBM and VBM' 
have a strong $p_{z}$ Se orbital character. 
However, we find that the so far neglected $p_{x}p_{y}$ orbitals play a key role for an accurate 
description of the orbital composition of the valence band maxima, as we discuss below.

Along the $\Gamma\rightarrow Z$ symmetry line, the $R_{3}$ symmetry is preserved. 
Thus, the $|\Lambda_{1/2},\pm1/2\rangle$ and $|\Lambda_{3/2},\pm3/2\rangle$ effective 
states do not mix. In contrast, in the $\Gamma\rightarrow L$ and $\Gamma\rightarrow F$ 
paths the $R_{3}$ symmetry is broken. This allows for the 
hybridization of $p_z$ atomic orbitals with $p_{x}$ and $p_{y}$ ones. We find that this 
hybridization can be rather large, as clearly shown by Figs.~\ref{Int1}b and \ref{Int1}c, 
where we present the Se orbital composition of the $J=1/2$ and $J=3/2$ bands along 
the Brillouin zone.

Since the valence band maxima do not belong to the $\Gamma\rightarrow Z$ symmetry line, 
their orbital composition is a superposition of all $p$ Se-atomic orbitals. 
As a consequence, a minimal Hamiltonian aiming to effectively describe  
VBM and VBM' needs to take into account the states associated with the $p_{x}$ and 
$p_{y}$ orbitals, instead of including just the states with $p_z$ character 
\cite{PhysRevB.82.045122,PhysRevB.84.115413}. 

To calculate the surface electronic structure in the presence of surface projected bulk states, 
we consider a tight-binding model with eight states, namely, 
the $|\text{Se}^{-}_{1/2},\pm 1/2\rangle$ and $|\text{Bi}^{+}_{1/2},\pm 1/2\rangle$ states 
responsible for the band inversion, 
and $|\text{Se}^{-}_{3/2},\pm 3/2\rangle$ and $|\text{Se}^{+}_{3/2},\pm 3/2\rangle$ that dominate 
the most energetic $J=3/2$ band.
Using this basis, we write the 8$\times$8 Hamiltonian:
\begin{equation}
\mathcal{H}(\boldsymbol{k}) = %\varepsilon(\boldsymbol{k}) + 
\left(\begin{array}{cc}
\mathcal{H}_{1/2}(\boldsymbol{k}) & \mathcal{H}_{\rm int}(\boldsymbol{k})\\
\mathcal{H}_{\rm int}^{\dagger}(\boldsymbol{k}) & \mathcal{H}_{3/2}(\boldsymbol{k})\\
\end{array}\right),
\label{eq1}
\end{equation}
where $\mathcal{H}_{1/2}(\boldsymbol{k})$ is the standard 4$\times$4 Hamiltonian discussed 
in the literature \cite{PhysRevB.82.045122,PhysRevB.84.115413}, that considers only 
$|{\rm Bi}^{+}_{1/2},\pm 1/2\rangle$ and $|{\rm Se}^{-}_{1/2},\pm 1/2\rangle$ states
\footnote{We note that Ref.~\onlinecite{PhysRevB.82.045122} presents an $8\times 8$ Hamiltonian,
which is slightly different from ours, but does not explore its consequences of the additional bands. 
The focus of this seminal paper is the study of $\mathcal{H}_{1/2}({\bm k})$. }
Our model introduces $\mathcal{H}_{3/2}(\boldsymbol{k})$, a 4$\times$4 Hamiltonian associated 
with the $|{\rm Se}^{-}_{3/2},\pm 3/2\rangle$ and $|{\rm Se}^{+}_{3/2},\pm 3/2\rangle$ states,
and $\mathcal{H}_{\rm int}(\boldsymbol{k})$ the corresponding coupling term.

For a given total angular momentum $J$ the matrix elements in $\mathcal{H}(\boldsymbol{k})$ read
\begin{equation}
[\mathcal{H}(\boldsymbol{k})]_{ii^\prime}=\varepsilon_{ii^\prime}(\boldsymbol{k})\delta_{ii^\prime}+
\sum_{\nu}\left(t^{ii^\prime}_{\boldsymbol{a}_{\nu}}e^{i\boldsymbol{k}\cdot\boldsymbol{a}_{\nu}}
 +t^{ii^\prime}_{\boldsymbol{b}_{\nu}}e^{i\boldsymbol{k}\cdot\boldsymbol{b}_{\nu}}\right),
\label{matrixelements}
\end{equation}
where the states are labeled by $i=(\Lambda,J,\tau,j_{z})$, $\varepsilon_{ii}(\boldsymbol{k})$ 
are on-site energy terms, and $t^{ii^\prime}_{\bm c}=\langle{\bm n},\Lambda^{\tau}_{J},j_{z}|H|
\boldsymbol{n}+{\bm c},\Lambda^{'\tau'}_{J'},j_{z}'\rangle$ are the corresponding 
nearest neighbor QL hopping terms, with $\boldsymbol{n}_\nu$ and $\tau$
indicating lattice site and orbital parity, respectively. Here ${\bm c} = {\bm a}_\nu$ or
${\bm b}_\nu$, where
$\pm\boldsymbol{a}_{\nu}$ stands for the 6 intra-layer nearest neighbor vectors of each 
triangular lattice, namely,  
${\bm a}_{1}=(a,0,0), \boldsymbol{a}_{2}=(-a/2,\sqrt{3}a/2,0), \boldsymbol{a}_{3}=(-a/2,-\sqrt{3}a/2,0)$,
while $\pm\boldsymbol{b}_{\nu}$ denotes the 6 inter-layer nearest neighbors vectors, 
${\bm b}_{1}=(0,\sqrt{3}a/3,c/3), {\bm b}_{2}=(-a/2,-\sqrt{3}a/6,c/3), {\bm b}_{3}=(a/2,-\sqrt{3}a/6,c/3)$
with $a = 4.14$ \AA~and $c = 28.70 $ \AA~\cite{Zhang2009}.

%-------------------------------------   FIGURE 2  -----------------------------------------------------
\begin{figure}
\includegraphics[width = 0.95\columnwidth]{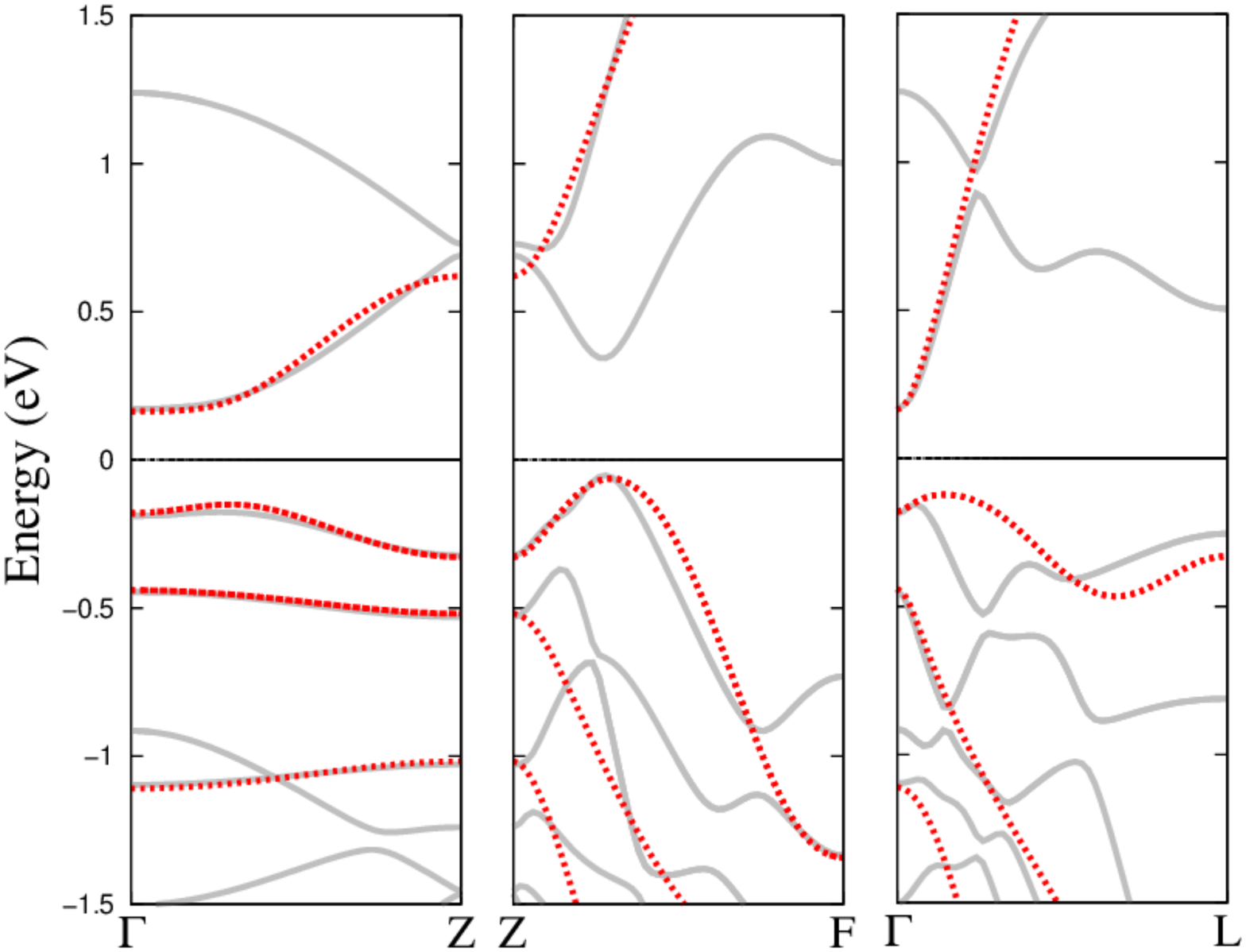}
\caption{(Color online) Comparison between the DFT (gray solid lines) and the tight-binding model (red dotted lines) 
bulk band structure of Bi$_{2}$Se$_{3}$.  }
\label{fig:TB_bulk}
\end{figure}
%---------------------------------   END FIGURE 2 ---------------------------------------------------

Exploring the system symmetries, we find constraints relating the nearest neighbors 
QL hopping terms $t^{ij}_{\bm c}$, thereby reducing the total number of possible 
hopping terms from 432 to 30 independent ones (see Appendix \ref{sec:appA}). 
The corresponding 30 tight-binding parameters are determined by fitting the tight-binding model
bulk band structure to the one calculated with DFT, shown in Fig~\ref{fig:TB_bulk}.
We present the complete Hamiltonian and provide more details on the fitting procedure 
in Appendix \ref{sec:appA}. 

The proposed Hamiltonian captures the low-energy {\it ab initio} band dispersion, even for
 $k$-points far from $\Gamma$, overcoming an intrinsic limitation of the ${\bm k}\cdot {\bm p}$ 
 models proposed in the literature to describe the band inversion at the $\Gamma$ point. 
We show in the Appendix \ref{sec:appA} how to reduce our model to a  ${\bm k}\cdot {\bm p}$ 
Hamiltonian by taking the approximation $k\rightarrow\Gamma$ and relating, for instance, the 
hopping terms $t^{ii^\prime}_{\boldsymbol{a}_{\nu}}$ and $t^{ii^\prime}_{\boldsymbol{b}_{\nu}}$ to 
the perturbation theory parameters of Ref.~\onlinecite{PhysRevB.82.045122}. The inclusion of 
additional bands does not affect the band inversion, for instance, the $J=3/2$ bands have much 
lower energies than the $J=1/2$ bands. 
%We now focus the discussion on the effect on the interaction between $J=3/2$ and $J=1/2$ states.

%%%%%%%%%%%%%%%%%%%%%%%%%%%%%%%%%%%%%%%%%%%%
\section{Thin films} 
\label{sec:thin_films}
%%%%%%%%%%%%%%%%%%%%%%%%%%%%%%%%%%%%%%%%%%%%

In this section we calculate the electronic band structure of rhombohedral TI thin films. 
We take the QLs parallel to the $xy$-plane and define the $z$-axis as the stacking direction.
The thickness of the films is given in terms of $N_{\rm QL}$, the number of stacked QLs.
The surface corresponds to the outermost QLs. The surface states correspond to the ones
spatially localized in these QLs.

We modify the bulk tight-binding Hamiltonian defined in Eq.~(\ref{eq1}) to account for a finite 
number of layers. The slab Hamiltonian consists of intra- and inter-layer terms, 
namely \cite{Ebihara2012885}
\begin{equation} 
\mathcal{H}_{\rm slab}=
%(k_{x},k_{y})
%\varepsilon_{slab}(k_{x}k_{y})+
\sum_{n=1}^{N_{\rm QL}}{c}^{\dagger}_{n}\mathcal{H}^{}_{0}{c}_{n}^{}+
\sum_{n=1}^{N_{{\rm QL}}-1}\left({c}^{\dagger}_{n}\mathcal{H}^{}_{z}{c}^{}_{n+1}+ \rm {H.c.}\right).
\label{Slab}
\end{equation}
The basis is given by $|n, k_{x},k_{y},\Lambda^{\tau}_{J},j_{z}\rangle$ with corresponding 
creation (annihilation) operators given in compact notation by $c^{\dagger}_{n}$ ($c_{n}^{}$). 
The intra-layer matrix elements read
\begin{equation}
[\mathcal{H}_{0}({\bm k})]_{ii^\prime}=\varepsilon_i({\bm k})\delta_{ii^\prime}+
\sum_{\nu=1}^{6}t^{ii^\prime}_{\boldsymbol{a}_{\nu}}e^{i\boldsymbol{k}\cdot\boldsymbol{a}_{\nu}}.
\label{eq:interlayer}
\end{equation}
The latter  are similar to those of Eq.~\eqref{matrixelements}, but restricted to two-dimensions, namely,
${\bm k}= (k_x, k_y)$.
In turn, the inter-layer term, 
\begin{equation}
[\mathcal{H}_{z}]_{ii^\prime}=\sum_{\nu}t^{ii^\prime}_{\boldsymbol{b}_{\nu}},
\label{eq:intralayer}
\end{equation}
provides the coupling between nearest neighbor QLs planes. 

It is well established that a bulk band inversion occurs between states dominated by 
$p_{z}$ Se and Bi atomic orbitals with different parities \cite{Zhang2009}. 
The four states $|\text{Se}^{-}_{1/2},\pm 1/2\rangle$ and 
$|\text{Bi}^{+}_{1/2},\pm 1/2\rangle$ form a good basis to describe the surface states 
at the $k$-points near the $\Gamma$ point \cite{PhysRevB.82.045122,PhysRevB.84.115413}. 
However, similarly to bulk systems, this reduced basis also fails to correctly describe the 
bulk states close in energy to the Dirac point in thin Bi$_2$Se$_3$ films.

%------------------------------------------ FIGURE 3 ------------------------------------------------------
\begin{figure}[h!]
\includegraphics[width = 0.9\linewidth]{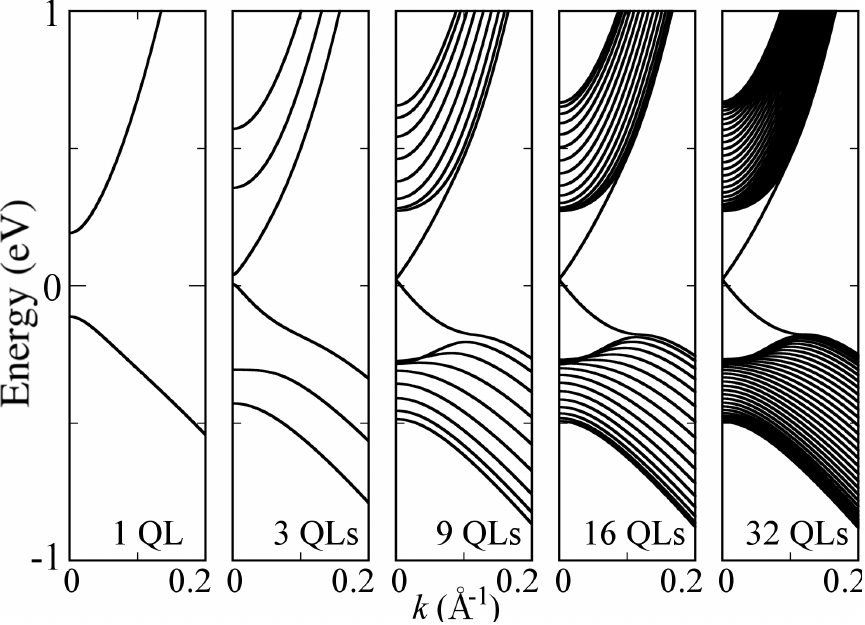}
\caption{Band structure along the $\Gamma\rightarrow M$ symmetry line,
 {without considering $J=3/2$-states}, for different film thicknesses of Bi$_{2}$Se$_{3}$.}
\label{fig:films_reducedH}
\end{figure}
%--------------------------------------- END  FIGURE 3 -------------------------------------------------

To better understand the importance of the $J=3/2$ states, let us first consider a thin film 
described by the Hamiltonian $\mathcal{H}_{1/2}({\boldsymbol{k}})$ projected out from 
$\mathcal{H}_{\rm slab}$.
Figure \ref{fig:films_reducedH} shows the finite size effects and how the band structure is 
modified by increasing $N_{\rm QL}$\cite{Ebihara2012885}.
For $N_{\rm QL}\geqslant 3$ one clearly observes the appearance of surface states and the 
formation of a Dirac cone. 
For $N_{\rm QL}\gg 1$ the bulk band gap is recovered. 
We stress that without the $J=3/2$ states, the model does not show VBM bulk states close 
to the Fermi level, as expected from the analysis of bulk band structure (see, for instance, Fig.~\ref{Int1}). 
Moreover, within this simple model the band structure close to the Dirac point along the 
$\Gamma\rightarrow K$ and $\Gamma\rightarrow M$ paths are identical, which is a rather unrealistic 
symmetry feature. 

%------------------------------------------ FIGURE 4 ------------------------------------------------------
\begin{figure}[h!]
\includegraphics[width= 0.95\linewidth]{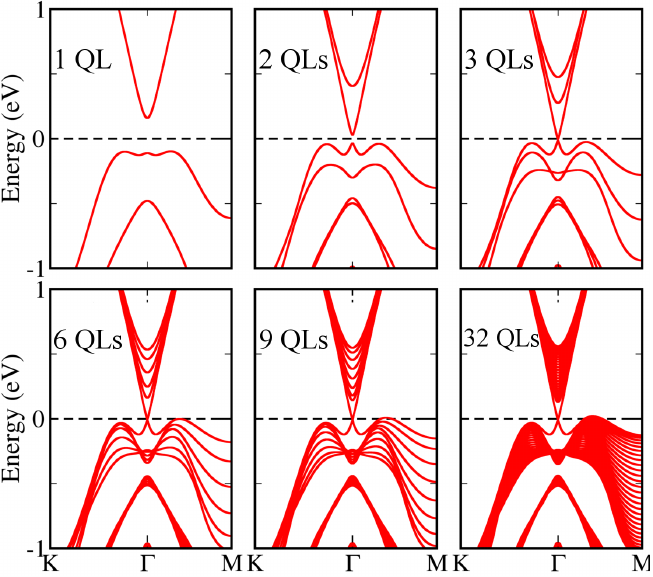}
\caption{(Color online) Band structure of a Bi$_{2}$Se$_{3}$ thin film for different thickness
values, $N_{\rm QL} = 1,2,3,6,9,$ and 32, using our 8$\times$8 tight-binding effective model.}
\label{fig:bs-varyingQLs}
\end{figure}
%---------------------------------------- END FIGURE 4 ----------------------------------------------------

The $J=3/2$ states modify significantly the electronic band structure. 
Figure \ref{fig:bs-varyingQLs} summarizes the results we obtain for the 8$\times$8 total 
effective Hamiltonian, Eq. \eqref{eq1}. Even for a few QLs, 
the shape of the surface band structure reproduces the qualitative behavior observed in the 
bulk LDA-DFT calculations. Figure \ref{fig:bs-varyingQLs} shows that as $N_{\rm QL}$ is 
increased, the Dirac cone is formed and bulk states appear in the vicinity of the 
Fermi level turning the system into a metal.

%%%%%%%%%%%%%%%%%%%%%%%%%%%%%%%%%%%%%%%%%%%%%
\section{Application: Bulk states engineering}
\label{sec:shift}
%%%%%%%%%%%%%%%%%%%%%%%%%%%%%%%%%%%%%%%%%%%%%

Several strategies have been proposed and used to suppress the scattering channels associated
with the continuous bulk states, like for instance, alloy stoichiometry \cite{ZhangJinsong2011,Arakane2012,PhysRevB.84.165311,Abdalla2015}, 
application of an external electric field\cite{PhysRevLett.105.266806}, stacking faults\cite{Seixas2013}, 
and strain\cite{LiuY2014,nanolett5b00553}. 

Let us now use the effective model put forward in the previous section to discuss some of these known 
strategies to shift the bulk band 
states away from the Dirac point energy, defined as $\varepsilon = 0$.

\begin{figure}[h!]
\includegraphics[width= 0.9\linewidth]{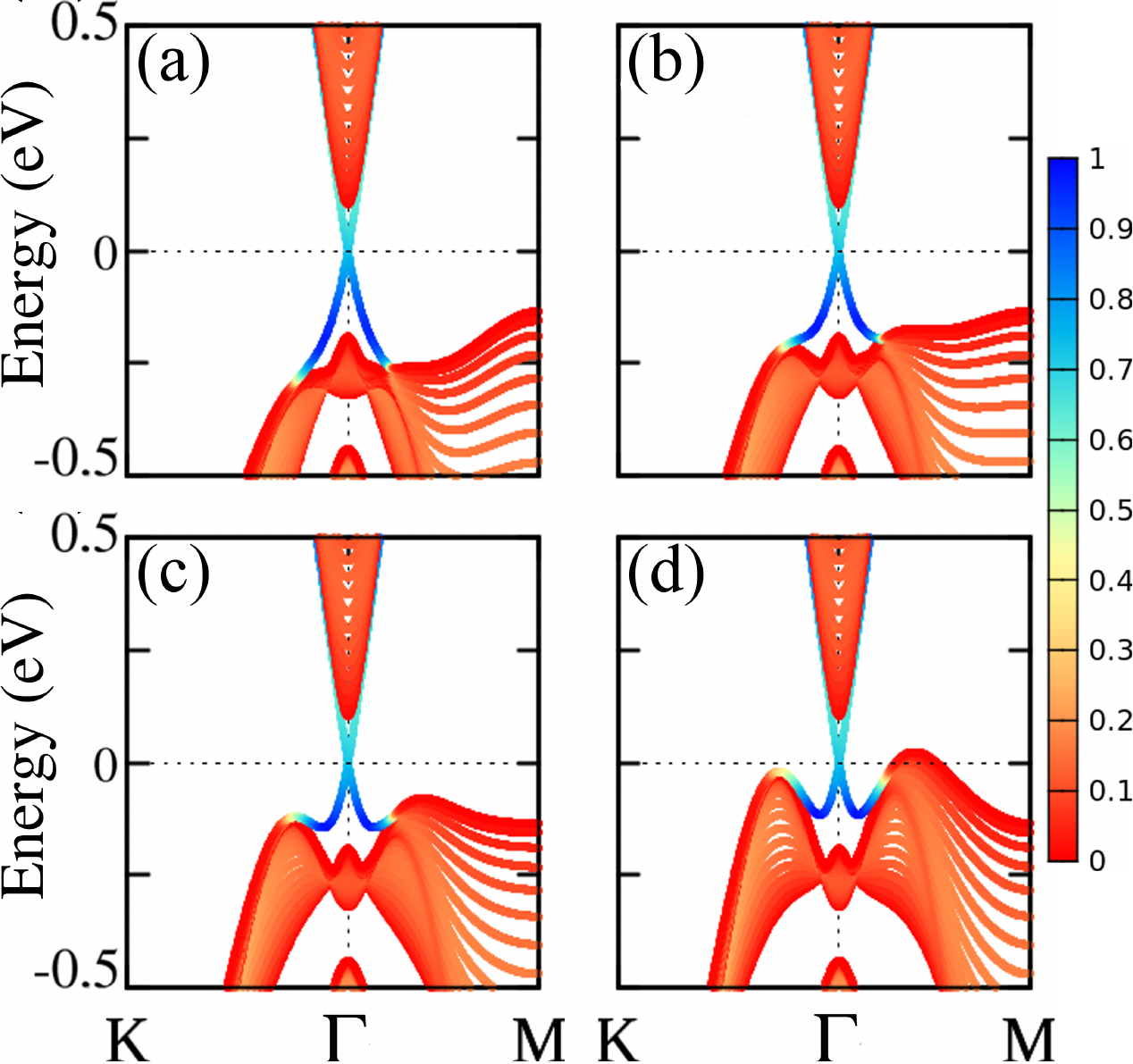}
\caption{(Color online) Surface band structure for $N_{\rm QL} =20$ calculated 
using the eight bands effective model with hopping term (repulsion parameter) (a) 
$t^{ij}_{b}= 1.2$ eV, (b) $t^{ij}_{b}= 1.5$ eV, 
(c) $t^{ij}_{b}= 1.8$ eV, and (d) $t^{ij}_{b}= 2.1$ eV, where $i=(\mbox{Se}_{1/2}^{-},\pm1/2)$,
 and 
$j=(\mbox{Se}_{3/2}^{+},\pm3/2)$. 
The color code stands for the magnitude of the projection of the orbitals at the outermost (surface) QLs.
Pure surface states are indicated by blue, whereas bulk states are depicted in red.}
\label{fig:Se-Se_hopping}
\end{figure}

Our analysis is based on the observation that the energy of the bulk states along the
$\Gamma\rightarrow M$ symmetry path depends very strongly on the in-plane interaction between 
$|{\rm Se}^{-}_{1/2},\pm 1/2\rangle$ and $|{\rm Se}^{+}_{3/2},\pm 3/2\rangle$ states. 
We find that by increasing the matrix elements associated with the mixture of 
the above states the bulk states are shifted up in energy, as shown by 
Fig.~\ref{fig:Se-Se_hopping}. 

%------------------------------------------ FIGURE 5 ------------------------------------------------------
%---------------------------------------- END FIGURE 5 ----------------------------------------------------

Hence, as previously proposed \cite{Abdalla2015}, one way to engineer the VBM and VBM$^\prime$ 
states is by substituting the Se atoms by chemical elements that do not spoil the topological properties 
of the material and reduce the interaction between $|{\rm Se}^{-}_{1/2},\pm 1/2\rangle$ and 
$|{\rm Se}^{+}_{3/2},\pm 3/2\rangle$ states. 
This effect can be described by a simple model in terms of the direct modification of the matrix element
$t^{ij}_{b}$ that mixes the $|{\rm Se}^{-}_{1/2},\pm 1/2\rangle$ and $|{\rm Se}^{+}_{3/2},\pm 3/2\rangle$ 
states. 
In fact, the band structures obtained for several values of $t^{ij}_{b}$ shown in Fig.~\ref{fig:Se-Se_hopping} 
qualitatively describe the first-principles calculations for Bi$_2$(Se$_{1-x}$S$_{x}$)$_{3}$ alloys \cite{Abdalla2015}.

%We propose that the inclusion of S atoms in the Bi$_2$Se$_3$-like materials is a suitable strategy for controlling the valence band maxima energies. 
%For instance, the VBM that is clearly observed in the surface band structure, left panel of 
%Fig.~\ref{fig:alloy}, is suppressed in the Bi$_2$(Se$_{1-x}$S$_{x}$)$_{3}$ alloy with the specific 
%concentration $x=0.1$, as shown in the \textit{ab-initio} simulations displayed by Fig.~\ref{fig:alloy}. 

%------------------------------------------ FIGURE 5 ------------------------------------------------------
%\begin{figure}
%\includegraphics[width=8.6cm]{Fig5_correction.pdf}
%\caption{(Color online) Bi$_{2}$(Se$_{0.9}$S$_{0.1}$)$_{3}$ alloys: 
%(left)  Surface band structure (red lines) for 6QLs contrasted with the bulk band structure 2D projection (grey lines) for pristine Bi$_{2}$Se$_{3}$. The valence band maximum VBM is indicated. (right) Suppression of the continuous bulk states for 6QL at the Fermi energy for the Bi$_{2}$(Se$_{0.9}$S$_{0.1}$)$_{3}$ alloy. }
%\label{fig:alloy}
%\end{figure}
%--------------------------------------- END FIGURE 5 ----------------------------------------------------

\begin{figure}
\includegraphics[width=8.6cm]{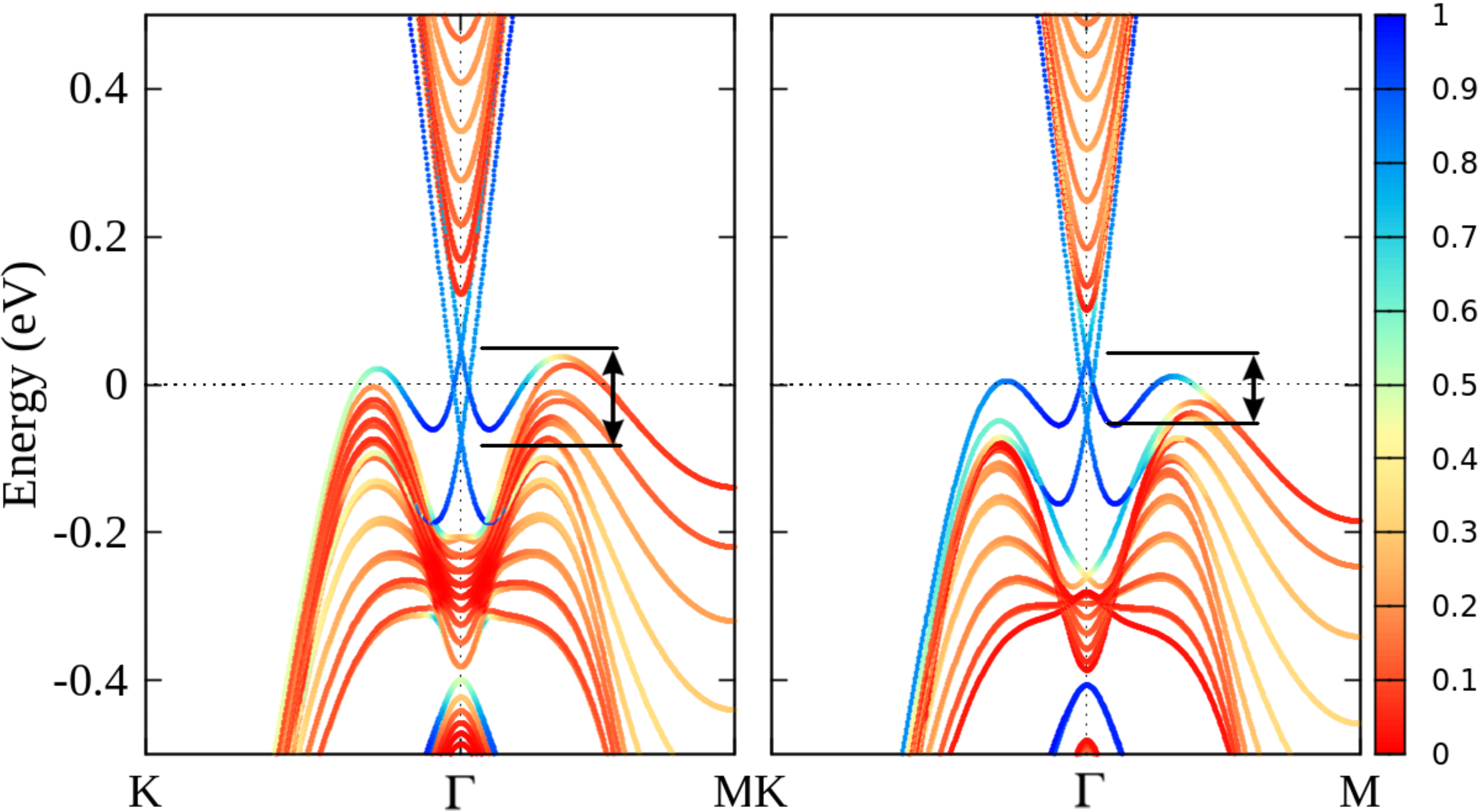} 
\caption{(Color online) Electric field (left) and stacking faults (right) effect of the 
band structure of a Bi$_2$Se$_3$ thin film of 9QLs. 
The splitting between the Dirac cones associated with different surfaces is 
represented by the arrow.
The color code quantifies the surface/bulk character of the electronic states, see caption of 
Fig. \ref{fig:Se-Se_hopping}.}
\label{Fig6}
\end{figure}

Alternatively, the double degenerate surface-state bands due to the presence of two [111] cleavage surfaces 
in a slab geometry can be removed by applying a perpendicular electric field $E_{0}\hat{z}$ 
\cite{PhysRevLett.105.266806}.
The Dirac cone associated with the surface at the highest potential energy can be shifted above the VBM, 
leading to a suppression of the scattering channels between the topologically protected 
metallic surface states and the bulk states. We describe this effect using our tight-binding effective model 
by modifying the on-site term $\varepsilon(\boldsymbol{k})\delta_{ij}$ in the inter-layer matrix elements
associated with each QL. As a result, Eq.~\eqref{eq:interlayer} becomes
\begin{equation}
[\mathcal{H}_{0}(k_{x}, k_{y})]_{n,ij}=\tilde{\varepsilon}_{n}(\boldsymbol{k})\delta_{ij}+
\sum_{\nu}t^{ij}_{\boldsymbol{a}_{\nu}}e^{i\boldsymbol{k}\cdot\boldsymbol{a}_{\nu}},
\end{equation}
where $\tilde{\varepsilon}_{n}(\boldsymbol{k})=\varepsilon(\boldsymbol{k})+ nceE_{0}/N_{QL}$, 
$n$ is the layer index, and $e$ is the electron charge. This simple approach captures the shift of the Dirac 
cone located at the surfaces corresponding to the QL with $n=N_{QL}$ and $n=0$. Figure \ref{Fig6}a show the
effect of an electric field of $E = 5\times 10^{-3}$ V/\AA~ on a thin film of $N_{\rm QL} = 9$. 

Another band engineering strategy has been suggested by \textit{ab-initio} atomistic investigations on the role 
played by extended  defects, like stacking faults, on the structural and electronic properties of 3D topological 
insulators~\cite{Seixas2013}.
In $R\overline{3}m$ structures the typical stacking is a ABCABC configuration, that is, each QL is rotated with
respect to its adjacent QL by 120$^{o}$. When a QL is ``removed" leading to a ACABCA, ABABCA, or ABCBCA 
stacking configuration, the defects is called an intrinsic stacking fault.
The inter-QLs distance decreases as a consequence of these stacking faults, 
making the Van der Waals inter-QLs interaction weaker and changing the on-site potential of the QLs
in which the structural defect is located~\cite{Seixas2013}.
Thus, it is relatively easy to account for this effect within our model, namely, we rewrite the on-site energy and 
the inter band interaction as $\varepsilon_{n}(\boldsymbol{k})-\delta\varepsilon_{0}$ and $t^{ij}_{\boldsymbol{b}_{\nu}}
-\delta t$. 
%Only for the rotated QL, $\delta\varepsilon_{0}$ and $\delta t$ are not null, as discused in Fig.~\ref{Fig6}%}.

Stacking faults nearby the surface layers of Bi$_2$Se$_3$ give rise to a 
positive energy shift of the bulk states with respect to their energy in a pristine system\cite{Seixas2013}.
This shift is typically about 75 meV. 
Thus, we obtain a qualitative description of the stacking faults effect by fitting $\delta\varepsilon_{0}$ and 
$\delta t$ to the DFT results only for the QLs with this structural defect, see Fig.~\ref{Fig6}.
Our simplified model and description allows for the study of thin films with a large number of QLs.

%modifying the on-site term $\varepsilon_{\boldsymbol{k}}$ and the intra-layer coupling only between the QLs with this structural defect, see Fig.~\ref{Fig6}.

%------------------------------------------ FIGURE 6 ------------------------------------------------------
%----------------------------------------END  FIGURE 6 --------------------------------------------------

%%%%%%%%%%%%%%%%%%%%%%%%%%%%%%%%%%%%%%%%%%%%
\section{Conclusions}
\label{sec:conclusions}
%%%%%%%%%%%%%%%%%%%%%%%%%%%%%%%%%%%%%%%%%%%%

We have revisited the band structure calculations of rhombohedral topological insulators, 
both bulk and thin films, and investigated the occurrence of bulk states at the Fermi level. 
Based on \textit{ab initio} calculations, we construct a simplified tight-binding model considering the 
states with angular momentum $J=1/2$ and $J=3/2$ and therefore, taking explicitly 
into account the $p_{x}p_{y}$ Se orbitals contributions. 

Our model shows that the energy of bulk states near the Dirac-point is associated 
with a band mixing, which is mainly ruled by the hopping term between $p_{z}$ and 
$p_{x}p_{y}$ states. 
The valence band maximum appears in the symmetry path in which the $R_{3}$ symmetry is broken.
In this situation, the $J=3/2$ states can mix with the $J=1/2$ ones. 

We illustrate the versatility of our tight-binding model by studying some strategies to eliminate 
and/or shift the bulk states away from the Fermi surface. We show that the band structures obtained 
using our simple model reproduce qualitatively very well computationally costly {\it ab initio} calculations found
in the literature.

In summary, we show that our simple effective model captures the main surface band structure features, 
allowing to explore strategies to perform a continuous bulk states engineering and opening the possibility 
to model disorder, which is ubiquitous in rhombohedral TIs and beyond the scope of {\it ab initio} calculations.

\acknowledgements
This work was supported by FAPESP (grant 2014/12357-3), CNPq (grant 308801/2015-6), and FAPERJ 
(grant E-26/202.917/2015).

\appendix

%%%%%%%%%%%%%%%%%%%%%%%%%%%%%%%%%%%%%%%%%%%%
\section{Full effective Hamiltonian and model parameters}
\label{sec:appA}
%%%%%%%%%%%%%%%%%%%%%%%%%%%%%%%%%%%%%%%%%%%%

As discussed in the main text, the form of proposed effective Hamiltonian is obtained by 
considering symmetry arguments only, which allows one to address the complete family of 
rhombohedral materials. In turn, the model parameters are determined by fitting the electronic 
properties obtained from a given first principle calculation. 

In this Appendix we discuss in detail the reasoning behind the construction of the model and
present explicit expressions for the matrix elements of the resulting effective Hamiltonian.
The Appendix presents also the model parameters for both Bi$_2$Se$_3$ and Bi$_2$Te$_3$ 
compounds.

Let us begin recalling that  the effective Hamiltonian $\mathcal{H}({\bm k})$, Eq.~\eqref{eq1} reads 
\begin{equation}
\mathcal{H}(\boldsymbol{k}) =  \left(\begin{array}{cc}
\mathcal{H}_{1/2}(\boldsymbol{k}) & \mathcal{H}_{\rm int}(\boldsymbol{k})\\
\mathcal{H}_{\rm int}^{\dagger}(\boldsymbol{k}) & \mathcal{H}_{3/2}(\boldsymbol{k})\\
\end{array}\right).
\end{equation}
The states with effective angular moment $J=3/2$ are combination of two representations 
of the double group $D_{3d}^{5}(R3m)$. Therefore, we consider the states with defined representation:
\begin{equation}
\left|\Lambda^{\pm},\Gamma_{4}\right\rangle=\frac{1}{\sqrt{2}}\left(|\Lambda^{\pm}_{3/2}, 3/2\rangle
+|\Lambda^{\pm}_{3/2}, -3/2\rangle\right),
\end{equation}
and 
\begin{equation}
\left|\Lambda^{\pm},\Gamma_{5}\right\rangle
=\frac{1}{\sqrt{2}}\left(|\Lambda^{\pm}_{3/2}, 3/2\rangle-|\Lambda^{\pm}_{3/2},  -3/2\rangle\right).
\end{equation}

The states $\{|\Lambda^{\tau}_{J},j_{z}\rangle\}$ are transformed by the symmetries operator as
\begin{enumerate}
\item \textit{Threefold rotation $R_3$}: 
\begin{align*}
\left|\Lambda^{\pm},\Gamma_{4,5}\right\rangle\rightarrow &-\left|\Lambda^{\pm},\Gamma_{4,5}\right\rangle,\\
\left|\Lambda^{\pm},\pm 1/2 \right\rangle\rightarrow &e^{\pm\frac{i\pi}{3}}\left|\Lambda^{\pm},\pm 1/2\right\rangle.
\end{align*}

\item \textit{Twofold rotation $R_2$}: \begin{align*}
\left|\Lambda^{\pm},\Gamma_{4}\right\rangle\rightarrow &\pm i\left|\Lambda^{\pm},\Gamma_{4}\right\rangle,\\
\left|\Lambda^{\pm},\Gamma_{5}\right\rangle\rightarrow &\mp i\left|\Lambda^{\pm},\Gamma_{5}\right\rangle,\\
\left|\Lambda^{+},\pm 1/2 \right\rangle\rightarrow & i\left|\Lambda^{+},\mp 1/2\right\rangle,\\
\left|\Lambda^{-},\pm 1/2 \right\rangle\rightarrow & -i\left|\Lambda^{-},\mp 1/2\right\rangle. 
\end{align*}

\item \textit{Inversion $\mathcal{P}$}: \begin{align*} 
\left|\Lambda^{\pm},\Gamma_{4,5}\right\rangle\rightarrow &\pm\left|\Lambda^{\pm},\Gamma_{4,5}\right\rangle,\\
\left|\Lambda^{\pm},\alpha\right\rangle\rightarrow &\pm\left|\Lambda^{\pm},\alpha\right\rangle\quad \mbox{with}\quad\alpha=\pm 1/2.
\end{align*}

\item \textit{Time reversal $\mathcal{T}$}: 
\begin{align*} 
\left|\Lambda^{\pm},\Gamma_{4,5}\right\rangle\rightarrow &-\left|\Lambda^{\pm},\Gamma_{5,4}\right\rangle,\\
\left|\Lambda,\pm 1/2\right\rangle\rightarrow&\pm\left|\Lambda,\mp 1/2\right\rangle. 
\end{align*}
\end{enumerate}

These symmetry transformations relate the hopping terms to each other, as shown for 
$t_{{\bm c}_\nu}^{11}=\langle 0, {\rm Bi}^{+}_{1/2}, 1/2 |H| {\bm c}_\nu, {\rm Bi}^{+}_{1/2}, +1/2 \rangle$ in Table~\ref{symmetry11}.

%----------------------------------------------- T A B L E  1 -----------------------------------
\begin{table}[h!]
\center
\caption{Symmetry operations on the hopping matrix element $t_{{\bm c}_\nu}^{11}=\langle 0, {\rm Bi}^{+}_{1/2}, +1/2 |H| 
{\bm c}_{\nu}, {\rm Bi}^{+}_{1/2}, +1/2 \rangle$, where ${\bm c}_\nu = {\bm a}_\nu$ or ${\bm b}_\nu$, and $\nu =1,2$ or 3.
(For completeness, we recall that $t_{{\bm c}_\nu}^{22}=\langle 0, {\rm Bi}^{+}_{1/2}, -1/2 |H| 
{\bm c}_{\nu}, {\rm Bi}^{+}_{1/2}, -1/2 \rangle$.)
%
%\CAIO{I changed the notation to be the same as in the main text. Please verify.}
}
\begin{tabular}{| c | c | c | c | c | c | c | }
\hline
               & $t_{a_{1},b_{1}}^{11}$  &  $t_{{a}_{2},b_{2}}^{11}$  &  $t_{{a}_{3},b_{3}}^{11}$  &  $t_{-a_{1},-b_{1}}^{11}$  &  $t_{-a_{2},-b_{2}}^{11}$  &  $t_{-a_{3},-b_{3}}^{11}$\\ 
\hline
$\mathcal{T}$  & $t_{a_{1},b_{1}}^{22*}$ &  $t_{a_{2},b_{2}}^{22*}$   &  $t_{a_{3},b_{3}}^{22*}$   &  $t_{-a_{1},-b_{1}}^{22*}$ &  $t_{-a_{2},-b_{2}}^{22*}$ &  $t_{-a_{3},-b_{3}}^{22*}$ \\
\hline
$\mathcal{P}$ & $t_{-a_{1},-b_{1}}^{11}$ &  $t_{-a_{2},-b_{2}}^{11}$   &  $t_{-a_{3},-b_{3}}^{11}$   &  $t_{a_{1},b_{1}}^{11}$   &  $t_{a_{2},b_{2}}^{11}$   &  $t_{a_{3},b_{3}}^{11}$   \\ 
\hline
$R_3$          & $t_{a_{3},b_{3}}^{11}$  &  $t_{a_{1},b_{1}}^{11}$    &  $t_{a_{2},b_{2}}^{11}$    &  $t_{-a_{3},-b_{3}}^{11}$  &  $t_{-a_{1},-b_{1}}^{11}$  &  $t_{-a_{2},-b_{2}}^{11}$  \\
\hline
$R_{2}$        & $t_{a_{1},b_{1}}^{22}$ &  $t_{a_{3},b_{3}}^{22}$   &  $t_{a_{2},b_{2}}^{22}$   &  $t_{-a_{1},-b_{1}}^{22}$   &  $t_{-a_{3},-b_{3}}^{22}$   &  $t_{-a_{2},-b_{2}}^{22}$   \\ 
\hline
\end{tabular}
\label{symmetry11}
\end{table}
%---------------------------------- E N D  T A B L E -----------------------------------------
	
These relations allow us to write the matrix elements in a simplified way. 
For instance, the matrix element $[\mathcal{H}(\boldsymbol{k})]_{11}$,~Eq.~\eqref{matrixelements}, 
is written as
\begin{equation}
[\mathcal{H}(\boldsymbol{k})]_{11}=\varepsilon_{11}(\boldsymbol{k})+ \alpha_{11}(\boldsymbol{k}),
\end{equation}
with
\begin{align} 
\alpha_{11}(\boldsymbol{k})=& \sum_{\nu=1}^{6}\left(t^{11}_{\boldsymbol{a}_{\nu}}e^{i\boldsymbol{k}\cdot\boldsymbol{a}_{\nu}}
 +t^{11}_{\boldsymbol{b}_{\nu}}e^{i\boldsymbol{k}\cdot\boldsymbol{b}_{\nu}}\right)
\nonumber\\
=& \sum_{\nu=1}^{3}\left(t^{11}_{\boldsymbol{a}_{\nu}}e^{i\boldsymbol{k}\cdot\boldsymbol{a}_{\nu}}
+t^{11}_{-\boldsymbol{a}_{\nu}}e^{-i\boldsymbol{k}\cdot\boldsymbol{a}_{\nu}}
\right.\nonumber\\
& \left.\;\;\;\;\;
+t^{11}_{\boldsymbol{b}_{\nu}}e^{i\boldsymbol{k}\cdot\boldsymbol{b}_{\nu}}
+t^{11}_{-\boldsymbol{b}_{\nu}}e^{-i\boldsymbol{k}\cdot\boldsymbol{b}_{\nu}}\right).
\end{align}
Using Table~\ref{symmetry11}, $\alpha_{11}(\boldsymbol{k})$ can be rewritten as 
\begin{align}
\alpha_{11}=& \sum_{\nu=1}^{3}\left[t^{11}_{\boldsymbol{a}_{\nu}}(e^{i\boldsymbol{k}\cdot\boldsymbol{a}_{\nu}}
+e^{-i\boldsymbol{k}\cdot\boldsymbol{a}_{\nu}})
+t^{11}_{\boldsymbol{b}_{\nu}}(e^{i\boldsymbol{k}\cdot\boldsymbol{b}_{\nu}}+e^{-i\boldsymbol{k}\cdot\boldsymbol{b}_{\nu}})\right]
\nonumber\\ 
=& 6\left(t^{11}_a\cos{\boldsymbol{k}\cdot\boldsymbol{a}_{\nu}}
+t^{11}_b\cos{\boldsymbol{k}\cdot\boldsymbol{b}}\right).
\end{align}
Time-reversal symmetry $\mathcal{T}$ and the two-fold rotation $R_{2}$ impose the relation 
$t_{{\bm a}_{1}}^{11}=t^{22*}_{{\bm a}_{1}}=t_{{\bm a}_{1}}^{22}$, which in turn requires 
$t_{{\bm a}_{1}}^{11}$ be real. 
A symmetry analysis, expanding Table~\ref{symmetry11} to other $ij$ values, shows 
that $t^{ij}_{{\bm a}_\nu} = t^{ ij}_a$
and $t^{ij}_{{\bm b}_\nu} = t^{ij}_b$. 

In the same way, we use the symmetry operations to calculate all terms for the Hamiltonian matrix 
elements describing rhombohedral TIs, which also imposes the sign and imaginary phases of the 
hopping terms, as presented below.
%
%\CAIO{From now on (see equations bellow and Table III, the hopping matrix elements do not depend on the vectors ${\bm a}_\nu$ and ${\bm b}_\nu$ as above, but rather on $a$ and $b$. Why? Symmetry considerations? Strange, since the symmetry operations in Table I give different results depending on $\nu$. It should be justified.}

The 4$\times$4 Hamiltonian $\mathcal{H}_{1/2}({\bm k})$, associated with the 
$|{\rm Se}^{-}_{1/2},\pm 1/2\rangle$ and $|{\rm Bi}^{+}_{1/2},\pm 1/2\rangle$ 
states, reads
\begin{equation}
\mathcal{H}_{1/2}(\boldsymbol{k}) =\left(\begin{array}{cccc}
\varepsilon_{1}+\alpha_{11} &      0                &i\alpha_{13}                &i\alpha_{14}\\
                       &\varepsilon_{1}+\alpha_{11} &-i\alpha_{14}^{*}           &i\alpha_{13}^{*}\\
\mbox{H.c.}            &                       &\varepsilon_{3}+\alpha_{33} & 0       \\
                       &                       &                       &\varepsilon_{3}+\alpha_{33	}\\
\end{array}\right),
\end{equation}
where the diagonal $\alpha_{ii}$ matrix elements are given by
\begin{equation}
\alpha_{ii}=2t_{a}^{ii}\sum_{\nu=1}^{3}\cos(\boldsymbol{k}\cdot\boldsymbol{a}_{\nu})+
2t_{b}^{ii}\sum_{\nu=1}^{3}\cos(\boldsymbol{k}\cdot\boldsymbol{b}_{\nu})
\end{equation}
while the off-diagonal ones read
\begin{equation}
\alpha_{13}=-i2t_{a}^{13}\sum_{j=1}^{3}\sin(\boldsymbol{k}\cdot\boldsymbol{a}_{j})+
2t_{b}^{13}\sum_{j=1}^{3}\sin(\boldsymbol{k}\cdot\boldsymbol{b}_{j})
\end{equation}
\begin{align}
\alpha_{14}&= 2t_{a}^{14}[\sin({\bm k}\cdot\boldsymbol{a}_{1})+ e^{-i2\pi/3}\sin(\boldsymbol{k}\cdot\boldsymbol{a}_{2}) \\ &
\nonumber
 +e^{-i4\pi/3}\sin(\boldsymbol{k}\cdot\boldsymbol{a}_{3})] \nonumber  +2t_{b}^{14}[e^{-i\pi/2}\sin(\boldsymbol{k}\cdot\boldsymbol{b}_{1}) \\&
 \nonumber
 +e^{i5\pi/6}\sin(\boldsymbol{k}\cdot\boldsymbol{b}_{2})+e^{i\pi/6}\sin(\boldsymbol{k}\cdot\boldsymbol{b}_{3})]. 
 \nonumber
\end{align}

The 4$\times$4 Hamiltonian, $H_{3/2}(\boldsymbol{k})$, 
associated with the $|{\rm Se}^{-}_{3/2},\pm 3/2\rangle$ and
$|{\rm Bi}^{+}_{3/2},\pm 3/2\rangle$ states is written as
\begin{equation}
\mathcal{H}_{3/2}(\boldsymbol{k}) =\left(\begin{array}{cccc}
\varepsilon_{5}+\beta_{55} &      0                & i\beta_{57}                & i\beta_{58}\\
                       &\varepsilon_{5}+\beta_{5} & -i\beta_{58}^{*}           & i\beta_{57}^{*}\\
\mbox{H.c.}            &                       &\varepsilon_{7}+\beta_{77} & 0       \\
                       &                       &                       &\varepsilon_{7}+\beta_{77}\\
\end{array}\right),
\end{equation}
where
\begin{equation}\beta_{ii}=2t_{a}^{ii}\sum_{j=1}^{3}\cos(\boldsymbol{k}\cdot\boldsymbol{a}_{j})+
2t_{b}^{ii}\sum_{j=1}^{3}\cos(\boldsymbol{k}\cdot\boldsymbol{b}_{j})
\end{equation}
\begin{equation}\beta_{57}=2t_{b}^{57}\sum_{j=1}^{3}\sin(\boldsymbol{k}\cdot\boldsymbol{b}_{j})
\end{equation}
\begin{equation}
\beta_{58}=2t_{a}^{58}\sum_{j=1}^{3}\sin(\boldsymbol{k}\cdot\boldsymbol{a}_{i}).
\end{equation}

The diagonal on-site energies $\varepsilon_i$ of the matrices $H_{1/2}(\boldsymbol{k})$ and 
$H_{3/2}(\boldsymbol{k})$  are given in Table \ref{table:epsilon}.

%\begin{equation}
%H_{\rm int}(\boldsymbol{k}) =\left(\begin{array}{cccc}
%\gamma^{a}_{15} +\gamma^{b}_{15} & \gamma^{a*}_{15}-\gamma^{b*}_{15}    & \gamma^{a}_{17} +\gamma^{b}_{17} & -\gamma^{a*}_{17}-\gamma^{b*}_{17}\\
%-\gamma^{a}_{15} +\gamma^{b}_{15} & \gamma^{a*}_{15} +\gamma^{b*}_{15}  & \gamma^{a}_{17} +\gamma^{b}_{17} & \gamma^{a}_{17}+\gamma^{b*}_{17}\\
%\gamma^{a}_{35} +\gamma^{b}_{35} &-\gamma^{a*}_{35}-\gamma^{b*}_{35}    & \gamma^{a}_{37} +\gamma^{b}_{37} & \gamma^{a*}_{37}-\gamma^{b*}_{37}\\
%\gamma^{a}_{35} +\gamma^{b}_{35} & \gamma^{a*}_{35}+\gamma^{b*}_{35}    & \gamma^{a}_{37} -\gamma^{b}_{37} & \gamma^{a*}_{37}+\gamma^{b*}_{37}\\
%\end{array}\right),
%\end{equation}
Finally, the interaction matrix $H_{\rm int}(\boldsymbol{k})$ is parametrized in block form as
\begin{equation}
\mathcal{H}_{\rm int}(\boldsymbol{k}) =\left(\begin{array}{cc}
{\bm \gamma}_{15} & {\bm \gamma}_{17}\\
{\bm \gamma}_{35} & {\bm \gamma}_{37}
\end{array}\right),
\end{equation}
where the ${\bm \gamma}$'s are 2$\times$ 2 matrices given by 
\begin{align}
{\bm \gamma}_{15}=& i\left(\begin{array}{cc}
\gamma^{a}_{15} +\gamma^{b}_{15} \;\;\;& \gamma^{a*}_{15}-\gamma^{b*}_{15}\\
-\gamma^{a}_{15} +\gamma^{b}_{15} \;\;\;& \gamma^{a*}_{15} +\gamma^{b*}_{15}\\
\end{array}\right),\\
{\bm \gamma}_{17}=&\left(\begin{array}{cc}
 \gamma^{a}_{17} +\gamma^{b}_{17}  \;\;\;& -\gamma^{a*}_{17}-\gamma^{b*}_{17}\\
 \gamma^{a}_{17} +\gamma^{b}_{17}  \;\;\;& \gamma^{a}_{17}+\gamma^{b*}_{17}\\
\end{array}\right),\\
{\bm \gamma}_{35}=&\left(\begin{array}{cc}
\gamma^{a}_{35} +\gamma^{b}_{35}  \;\;\;&-\gamma^{a*}_{35}-\gamma^{b*}_{35}\\
\gamma^{a}_{35} +\gamma^{b}_{35}  \;\;\;& \gamma^{a*}_{35}+\gamma^{b*}_{35}\\
\end{array}\right),\\
{\bm \gamma}_{37}=& i\left(\begin{array}{cc}
 \gamma^{a}_{37} +\gamma^{b}_{37}  \;\;\;& \gamma^{a*}_{37}-\gamma^{b*}_{37}\\
 \gamma^{a}_{37} -\gamma^{b}_{37}  \;\;\;& \gamma^{a*}_{37}+\gamma^{b*}_{37}\\
\end{array}\right).
\end{align}

Let us define
\begin{equation}
\Phi_c({\bm k}) = \sum_{\nu=1}^3 (-1)^{\nu-1} e^{-i(\nu-1)\pi/3}\sin(\boldsymbol{k}\cdot{\bm c}_{\nu})
\end{equation}
and
\begin{equation}
\Psi_c({\bm k}) = \sum_{\nu=1}^3 (-1)^{\nu-1} e^{-i(\nu-1)\pi/3}\cos(\boldsymbol{k}\cdot{\bm c}_{\nu}),
\end{equation}
where $c = a$ or $b$, to write 
\begin{align} 
\gamma_{15}^{a(b)} = & 2t_{a(b)}^{15} \Phi^{}_{a(b)}({\bm k}) ,
%\nonumber
\\
\gamma_{35}^{a(b)} = & 2t_{a(b)}^{35} \Psi^{}_{a(b)}({\bm k}),
%\nonumber
\\
\gamma_{17}^{a(b)} = & 2t_{a(b)}^{17} \Psi^{}_{a(b)}({\bm k}) ,
%\nonumber
\\
\gamma_{37}^{a(b)} = & 2t_{a(b)}^{37} \Phi^{}_{a(b)}({\bm k}) .
\end{align}

Symmetry considerations allow us to reduce the number of the model parameters to 30 independent ones. 
The latter are determined by a least-square fitting the bulk band structure obtained from the DFT calculation 
described in Sec. \ref{sec:tight-binding} for Bi$_2$Se$_3$ and Bi$_2$Te$_3$ rhombohedral materials.
The obtained on-site matrix elements are given in Table \ref{table:epsilon}, while the hopping matrix elements 
are shown in Table \ref{table:hopping}.

%-----------------------------  T A B L E   2 -----------------------------------
\begin{table}[h!]
\centering
\caption{On-site energies $\varepsilon_i$ (in eV). 
}
\begin{tabular}{ccccc}
\hline
\quad &\quad $\varepsilon_{1}$ \quad & $ \quad \varepsilon_{3}$ \quad & 
\quad $\varepsilon_{5}$ \quad & $ \quad \varepsilon_{7}$ \quad\\
\hline
Bi$_2$Se$_3$ \quad\quad & 1.602  \quad &  \quad -1.374  \quad&  \quad -1.050  \quad&  \quad -2.100 \\
Bi$_2$Te$_3$ \quad\quad & 0.805  \quad &  \quad -0.572  \quad&  \quad -0.9304  \quad&  \quad -1.900 \\
\hline
\end{tabular}
\label{table:epsilon}
\end{table}
%----------------------------------------------------------------------------------- 

%--------------------------------------- T A B L E   3  -----------------------------------------
\begin{table}[h!]
\centering
\caption{Non-zero hopping matrix elements $t_{c}^{ij}$ with $c=a,b$ in eV. The superscripts $ij$ 
listed by the first column correspond to all (symmetry) independent non vanishing hopping terms. 
%The zeros correspond to matrix elements smaller than $10^{-3}$~eV.
}
\begin{tabular}{ccccc}
\hline 
 &  & \hspace{-2cm} Bi$_{2}$Se$_{3}$ &  & \hspace{-2cm} Bi$_{2}$Te$_{3}$ \\
\hline
$ij$ \quad & \quad $t_b^{ij}$ (eV)  \quad & \quad $t_a^{ij}$(eV) \quad & \quad $t_b^{ij}$ (eV) \quad & \quad $t_a^{ij}$ (eV)\quad \\
\hline
$11$ \quad & $-0.067 \quad $ & $-0.240 $ & $-0.027$  \quad & $-0.130$ \\
$33$ \quad& $ 0.040 \quad $ & $ 0.211$   & $0.015$  \quad & $0.120$ \\
$55$ \quad& $ 0.0066 \quad $ & $ 0.095$ & $0.007$  \quad & $0.095$ \\
$77$ \quad& $-0.0097 \quad$ & $ 0.181$& $-0.012$  \quad & $0.171$ \\
%$12$ \quad& $ 0.000     \quad$ & $ 0.000 $\\
$13$ \quad& $ 0.045     \quad$ & $ 0.210$ & $-0.025$  \quad & $0.210$ \\
$14$ \quad& $i0.190 \quad$ & $-0.170$    & $i0.210$  \quad & $-0.270$ \\
$15$ \quad& $ 0.008 \quad$ & $ 0.100$    & $0.012$  \quad & $0.171$ \\
%$16$ \quad& $ 0.008 \quad$ & $ 0.100 $\\
$17$ \quad& $-0.008 \quad$ & $-0.120 + i0.006$ & $-0.012$  \quad & $-0.140 + i0.008$ \\
%$18$ \quad& $-0.008 \quad$ & $-0.120 + i0.006$\\
$35$ \quad& $-0.082 \quad$ & $0.152$   & $-0.093$  \quad & $0.092$\\
$37$ \quad& $-0.090 \quad$ & $0.210 $    & $-0.110$  \quad & $0.190$\\
$57$ \quad& $ < 10^{-3} \quad$ & $0.005$     & $< 10^{-3}$  \quad & $0.009$\\ 
$58$ \quad& $ 0.008 \quad$ & $< 10^{-3}$     & $0.012$  \quad & $< 10^{-3}$\\
\hline
\end{tabular}
\label{table:hopping}
\end{table}

The important parameters for the TI nature of the material are contained in the $\mathcal{H}_{1/2}$ Hamiltonian.
The role of the mass term (on-site term) in the band inversion is very well established in the literature, as well as all remaining matrix elements in $\mathcal{H}_{1/2}$~\cite{PhysRevB.84.115413,PhysRevB.82.045122}. 
The novelty here are the additional states that correctly account for surface projected bulk states, in which 
we have focused our discussion and are represented by $\mathcal{H}_{3/2}$. 
As discussed in Fig. \ref{Int1}b and Fig. \ref{Int1}c, we do not use an energy criterion, but
rather the total angular momentum and atomic orbitals projection to select the suitable basis
to describe the band interaction giving the shift in the bulk states. 
For instance, in Ref.~\onlinecite{PhysRevB.82.045122} the basis is \{$|\text{Se}^{-}_{1/2},\pm 1/2\rangle$, 
$|\text{Bi}^{+}_{1/2},\pm 1/2\rangle$, $|\text{Se}^{-}_{3/2},\pm 3/2\rangle$, and $|\text{Se}^{+}_{3/2},\pm 1/2\rangle$\}.
In our work we use \{$|\text{Se}^{-}_{1/2},\pm 1/2\rangle$ ,$|\text{Bi}^{+}_{1/2},\pm 1/2\rangle$,
$|\text{Se}^{-}_{3/2},\pm 3/2\rangle$, and $|\text{Se}^{+}_{3/2},\pm 3/2\rangle$\}. 
It is possible to compare the Hamiltonian matrix elements in Ref.~\onlinecite{PhysRevB.82.045122} with the 
ones obtained in this work only for the common elements, as shown in Table~\ref{table:hopping_Ref22}.

%--------------------------------------- T A B L E   3  -----------------------------------------
\begin{table}[h!]
\centering
\caption{Relation between the ${\bm k}\cdot {\bm p}$ perturbation theory parameters reported in 
Ref.~\onlinecite{PhysRevB.82.045122} with the hopping matrix elements.
}
\begin{tabular}{cc}
\hline 
 ${\bm k}\cdot {\bm p}$ parameters \qquad\quad & tight-binding parameters \quad \\
\hline
$F_{i(ij)}$ \quad & $-(a^{2}/2)(3t^{ii}_{a}+t^{ii}_{b})$ \\
$K_{i(ij)}$ \quad & $-3c^{2}t^{ii}_{b}$ \\
$Q_{1}$ \quad & $6ct^{13}_{b}$ \\
$P_{1}$ \quad & $a(3t^{14}_{a}-i\sqrt{3}t^{14}_{b})$ \\
$P_{2}=Q_{2}$ \quad & $a(3t^{15}_{a}+i\sqrt{3}t^{15}_{b})$ \\
$\overline{P}_{3}=\overline{Q}_{3}$ \quad & $a(3t^{17}_{a}+i\sqrt{3}t^{17}_{b})$ \\
$U_{35}$ \quad & $(a^{2}/4)(\frac{3}{4}t^{35}_{a}+t^{35}_{b})$ \\
$V_{35}$ \quad & $(i\sqrt{3}act^{35}_{b})$ \\
$\overline{U}_{37}$ \quad & $-(a^{2}/2)(3t^{37}_{a}+t^{37}_{b})$ \\
$\overline{V}_{37}$ \quad & $-3c^{2}t^{37}_{b}$ \\
\hline
\end{tabular}
\label{table:hopping_Ref22}
\end{table}

%----------------------------------------------------------------------------------------
%	BIBLIOGRAPHY
%----------------------------------------------------------------------------------------
\bibliography{3DTI} 

\begin{thebibliography}{37}%
\makeatletter
\providecommand \@ifxundefined [1]{%
 \@ifx{#1\undefined}
}%
\providecommand \@ifnum [1]{%
 \ifnum #1\expandafter \@firstoftwo
 \else \expandafter \@secondoftwo
 \fi
}%
\providecommand \@ifx [1]{%
 \ifx #1\expandafter \@firstoftwo
 \else \expandafter \@secondoftwo
 \fi
}%
\providecommand \natexlab [1]{#1}%
\providecommand \enquote  [1]{``#1''}%
\providecommand \bibnamefont  [1]{#1}%
\providecommand \bibfnamefont [1]{#1}%
\providecommand \citenamefont [1]{#1}%
\providecommand \href@noop [0]{\@secondoftwo}%
\providecommand \href [0]{\begingroup \@sanitize@url \@href}%
\providecommand \@href[1]{\@@startlink{#1}\@@href}%
\providecommand \@@href[1]{\endgroup#1\@@endlink}%
\providecommand \@sanitize@url [0]{\catcode `\\12\catcode `\$12\catcode
  `\&12\catcode `\#12\catcode `\^12\catcode `\_12\catcode `\%12\relax}%
\providecommand \@@startlink[1]{}%
\providecommand \@@endlink[0]{}%
\providecommand \url  [0]{\begingroup\@sanitize@url \@url }%
\providecommand \@url [1]{\endgroup\@href {#1}{\urlprefix }}%
\providecommand \urlprefix  [0]{URL }%
\providecommand \Eprint [0]{\href }%
\providecommand \doibase [0]{http://dx.doi.org/}%
\providecommand \selectlanguage [0]{\@gobble}%
\providecommand \bibinfo  [0]{\@secondoftwo}%
\providecommand \bibfield  [0]{\@secondoftwo}%
\providecommand \translation [1]{[#1]}%
\providecommand \BibitemOpen [0]{}%
\providecommand \bibitemStop [0]{}%
\providecommand \bibitemNoStop [0]{.\EOS\space}%
\providecommand \EOS [0]{\spacefactor3000\relax}%
\providecommand \BibitemShut  [1]{\csname bibitem#1\endcsname}%
\let\auto@bib@innerbib\@empty
%</preamble>
\bibitem [{\citenamefont {Hasan}\ and\ \citenamefont
  {Moore}(2011)}]{annurev-conmatphys-062910-140432}%
  \BibitemOpen
  \bibfield  {author} {\bibinfo {author} {\bibfnamefont {M.~Z.}\ \bibnamefont
  {Hasan}}\ and\ \bibinfo {author} {\bibfnamefont {J.~E.}\ \bibnamefont
  {Moore}},\ }\href {\doibase 10.1146/annurev-conmatphys-062910-140432}
  {\bibfield  {journal} {\bibinfo  {journal} {Annu. Rev. Condens. Matter
  Phys.}\ }\textbf {\bibinfo {volume} {2}},\ \bibinfo {pages} {55} (\bibinfo
  {year} {2011})}\BibitemShut {NoStop}%
\bibitem [{\citenamefont {Qi}\ and\ \citenamefont
  {Zhang}(2011)}]{RevModPhys.83.1057}%
  \BibitemOpen
  \bibfield  {author} {\bibinfo {author} {\bibfnamefont {X.-L.}\ \bibnamefont
  {Qi}}\ and\ \bibinfo {author} {\bibfnamefont {S.-C.}\ \bibnamefont {Zhang}},\
  }\href {\doibase 10.1103/RevModPhys.83.1057} {\bibfield  {journal} {\bibinfo
  {journal} {Rev. Mod. Phys.}\ }\textbf {\bibinfo {volume} {83}},\ \bibinfo
  {pages} {1057} (\bibinfo {year} {2011})}\BibitemShut {NoStop}%
\bibitem [{\citenamefont {Hasan}\ and\ \citenamefont
  {Kane}(2010)}]{RevModPhys.82.3045}%
  \BibitemOpen
  \bibfield  {author} {\bibinfo {author} {\bibfnamefont {M.~Z.}\ \bibnamefont
  {Hasan}}\ and\ \bibinfo {author} {\bibfnamefont {C.~L.}\ \bibnamefont
  {Kane}},\ }\href {\doibase 10.1103/RevModPhys.82.3045} {\bibfield  {journal}
  {\bibinfo  {journal} {Rev. Mod. Phys.}\ }\textbf {\bibinfo {volume} {82}},\
  \bibinfo {pages} {3045} (\bibinfo {year} {2010})}\BibitemShut {NoStop}%
\bibitem [{\citenamefont {Maciejko}\ \emph {et~al.}(2010)\citenamefont
  {Maciejko}, \citenamefont {Qi}, \citenamefont {Drew},\ and\ \citenamefont
  {Zhang}}]{PhysRevLett.105.166803}%
  \BibitemOpen
  \bibfield  {author} {\bibinfo {author} {\bibfnamefont {J.}~\bibnamefont
  {Maciejko}}, \bibinfo {author} {\bibfnamefont {X.-L.}\ \bibnamefont {Qi}},
  \bibinfo {author} {\bibfnamefont {H.~D.}\ \bibnamefont {Drew}}, \ and\
  \bibinfo {author} {\bibfnamefont {S.-C.}\ \bibnamefont {Zhang}},\ }\href
  {\doibase 10.1103/PhysRevLett.105.166803} {\bibfield  {journal} {\bibinfo
  {journal} {Phys. Rev. Lett.}\ }\textbf {\bibinfo {volume} {105}},\ \bibinfo
  {pages} {166803} (\bibinfo {year} {2010})}\BibitemShut {NoStop}%
\bibitem [{\citenamefont {Zhang}\ \emph {et~al.}(2009)\citenamefont {Zhang},
  \citenamefont {Liu}, \citenamefont {Qi}, \citenamefont {Dai}, \citenamefont
  {Fang},\ and\ \citenamefont {Zhang}}]{Zhang2009}%
  \BibitemOpen
  \bibfield  {author} {\bibinfo {author} {\bibfnamefont {H.}~\bibnamefont
  {Zhang}}, \bibinfo {author} {\bibfnamefont {C.-X.}\ \bibnamefont {Liu}},
  \bibinfo {author} {\bibfnamefont {X.-L.}\ \bibnamefont {Qi}}, \bibinfo
  {author} {\bibfnamefont {X.}~\bibnamefont {Dai}}, \bibinfo {author}
  {\bibfnamefont {Z.}~\bibnamefont {Fang}}, \ and\ \bibinfo {author}
  {\bibfnamefont {S.-C.}\ \bibnamefont {Zhang}},\ }\href
  {http://dx.doi.org/10.1038/nphys1270} {\bibfield  {journal} {\bibinfo
  {journal} {Nat. Phys.}\ }\textbf {\bibinfo {volume} {5}},\ \bibinfo {pages}
  {438} (\bibinfo {year} {2009})}\BibitemShut {NoStop}%
\bibitem [{\citenamefont {Mera~Acosta}\ \emph {et~al.}(2016)\citenamefont
  {Mera~Acosta}, \citenamefont {Babilonia}, \citenamefont {Abdalla},\ and\
  \citenamefont {Fazzio}}]{PhysRevB.94.041302}%
  \BibitemOpen
  \bibfield  {author} {\bibinfo {author} {\bibfnamefont {C.}~\bibnamefont
  {Mera~Acosta}}, \bibinfo {author} {\bibfnamefont {O.}~\bibnamefont
  {Babilonia}}, \bibinfo {author} {\bibfnamefont {L.}~\bibnamefont {Abdalla}},
  \ and\ \bibinfo {author} {\bibfnamefont {A.}~\bibnamefont {Fazzio}},\ }\href
  {\doibase 10.1103/PhysRevB.94.041302} {\bibfield  {journal} {\bibinfo
  {journal} {Phys. Rev. B}\ }\textbf {\bibinfo {volume} {94}},\ \bibinfo
  {pages} {041302} (\bibinfo {year} {2016})}\BibitemShut {NoStop}%
\bibitem [{\citenamefont {Eremeev}\ \emph {et~al.}(2012)\citenamefont
  {Eremeev}, \citenamefont {Landolt}, \citenamefont {Menshchikova},
  \citenamefont {Slomski}, \citenamefont {Koroteev}, \citenamefont {Aliev},
  \citenamefont {Babanly}, \citenamefont {Henk}, \citenamefont {Ernst},
  \citenamefont {Patthey}, \citenamefont {Eich}, \citenamefont {Khajetoorians},
  \citenamefont {Hagemeister}, \citenamefont {Pietzsch}, \citenamefont {Wiebe},
  \citenamefont {Wiesendanger}, \citenamefont {Echenique}, \citenamefont
  {Tsirkin}, \citenamefont {Amiraslanov}, \citenamefont {Dil},\ and\
  \citenamefont {Chulkov}}]{Eremeev2012}%
  \BibitemOpen
  \bibfield  {author} {\bibinfo {author} {\bibfnamefont {S.~V.}\ \bibnamefont
  {Eremeev}}, \bibinfo {author} {\bibfnamefont {G.}~\bibnamefont {Landolt}},
  \bibinfo {author} {\bibfnamefont {T.~V.}\ \bibnamefont {Menshchikova}},
  \bibinfo {author} {\bibfnamefont {B.}~\bibnamefont {Slomski}}, \bibinfo
  {author} {\bibfnamefont {Y.~M.}\ \bibnamefont {Koroteev}}, \bibinfo {author}
  {\bibfnamefont {Z.~S.}\ \bibnamefont {Aliev}}, \bibinfo {author}
  {\bibfnamefont {M.~B.}\ \bibnamefont {Babanly}}, \bibinfo {author}
  {\bibfnamefont {J.}~\bibnamefont {Henk}}, \bibinfo {author} {\bibfnamefont
  {A.}~\bibnamefont {Ernst}}, \bibinfo {author} {\bibfnamefont
  {L.}~\bibnamefont {Patthey}}, \bibinfo {author} {\bibfnamefont
  {A.}~\bibnamefont {Eich}}, \bibinfo {author} {\bibfnamefont {A.~A.}\
  \bibnamefont {Khajetoorians}}, \bibinfo {author} {\bibfnamefont
  {J.}~\bibnamefont {Hagemeister}}, \bibinfo {author} {\bibfnamefont
  {O.}~\bibnamefont {Pietzsch}}, \bibinfo {author} {\bibfnamefont
  {J.}~\bibnamefont {Wiebe}}, \bibinfo {author} {\bibfnamefont
  {R.}~\bibnamefont {Wiesendanger}}, \bibinfo {author} {\bibfnamefont {P.~M.}\
  \bibnamefont {Echenique}}, \bibinfo {author} {\bibfnamefont {S.~S.}\
  \bibnamefont {Tsirkin}}, \bibinfo {author} {\bibfnamefont {I.~R.}\
  \bibnamefont {Amiraslanov}}, \bibinfo {author} {\bibfnamefont {J.~H.}\
  \bibnamefont {Dil}}, \ and\ \bibinfo {author} {\bibfnamefont {E.~V.}\
  \bibnamefont {Chulkov}},\ }\href {http://dx.doi.org/10.1038/ncomms1638
  http://www.nature.com/ncomms/journal/v3/n1/suppinfo/ncomms1638\_S1.html}
  {\bibfield  {journal} {\bibinfo  {journal} {Nat. Commun.}\ }\textbf {\bibinfo
  {volume} {3}},\ \bibinfo {pages} {635} (\bibinfo {year} {2012})}\BibitemShut
  {NoStop}%
\bibitem [{\citenamefont {Liu}\ \emph {et~al.}(2010)\citenamefont {Liu},
  \citenamefont {Qi}, \citenamefont {Zhang}, \citenamefont {Dai}, \citenamefont
  {Fang},\ and\ \citenamefont {Zhang}}]{PhysRevB.82.045122}%
  \BibitemOpen
  \bibfield  {author} {\bibinfo {author} {\bibfnamefont {C.~X.}\ \bibnamefont
  {Liu}}, \bibinfo {author} {\bibfnamefont {X.~L.}\ \bibnamefont {Qi}},
  \bibinfo {author} {\bibfnamefont {H.~J.}\ \bibnamefont {Zhang}}, \bibinfo
  {author} {\bibfnamefont {X.}~\bibnamefont {Dai}}, \bibinfo {author}
  {\bibfnamefont {Z.}~\bibnamefont {Fang}}, \ and\ \bibinfo {author}
  {\bibfnamefont {S.~C.}\ \bibnamefont {Zhang}},\ }\href {\doibase
  10.1103/PhysRevB.82.045122} {\bibfield  {journal} {\bibinfo  {journal} {Phys.
  Rev. B}\ }\textbf {\bibinfo {volume} {82}},\ \bibinfo {pages} {045122}
  (\bibinfo {year} {2010})}\BibitemShut {NoStop}%
\bibitem [{\citenamefont {Yang}\ \emph {et~al.}(2012)\citenamefont {Yang},
  \citenamefont {Setyawan}, \citenamefont {Wang}, \citenamefont {{Buongiorno
  Nardelli}},\ and\ \citenamefont {Curtarolo}}]{Yang2012}%
  \BibitemOpen
  \bibfield  {author} {\bibinfo {author} {\bibfnamefont {K.}~\bibnamefont
  {Yang}}, \bibinfo {author} {\bibfnamefont {W.}~\bibnamefont {Setyawan}},
  \bibinfo {author} {\bibfnamefont {S.}~\bibnamefont {Wang}}, \bibinfo {author}
  {\bibfnamefont {M.}~\bibnamefont {{Buongiorno Nardelli}}}, \ and\ \bibinfo
  {author} {\bibfnamefont {S.}~\bibnamefont {Curtarolo}},\ }\href {\doibase
  10.1038/nmat3332} {\bibfield  {journal} {\bibinfo  {journal} {Nat. Mater.}\
  }\textbf {\bibinfo {volume} {11}},\ \bibinfo {pages} {614} (\bibinfo {year}
  {2012})}\BibitemShut {NoStop}%
\bibitem [{\citenamefont {Xia}\ \emph {et~al.}(2009)\citenamefont {Xia},
  \citenamefont {Qian}, \citenamefont {Hsieh}, \citenamefont {Wray},
  \citenamefont {Pal}, \citenamefont {Lin}, \citenamefont {Bansil},
  \citenamefont {Grauer}, \citenamefont {Hor}, \citenamefont {Cava},\ and\
  \citenamefont {Hasan}}]{Xia2009}%
  \BibitemOpen
  \bibfield  {author} {\bibinfo {author} {\bibfnamefont {Y.}~\bibnamefont
  {Xia}}, \bibinfo {author} {\bibfnamefont {D.}~\bibnamefont {Qian}}, \bibinfo
  {author} {\bibfnamefont {D.}~\bibnamefont {Hsieh}}, \bibinfo {author}
  {\bibfnamefont {L.}~\bibnamefont {Wray}}, \bibinfo {author} {\bibfnamefont
  {A.}~\bibnamefont {Pal}}, \bibinfo {author} {\bibfnamefont {H.}~\bibnamefont
  {Lin}}, \bibinfo {author} {\bibfnamefont {A.}~\bibnamefont {Bansil}},
  \bibinfo {author} {\bibfnamefont {D.}~\bibnamefont {Grauer}}, \bibinfo
  {author} {\bibfnamefont {Y.~S.}\ \bibnamefont {Hor}}, \bibinfo {author}
  {\bibfnamefont {R.~J.}\ \bibnamefont {Cava}}, \ and\ \bibinfo {author}
  {\bibfnamefont {M.~Z.}\ \bibnamefont {Hasan}},\ }\href
  {http://dx.doi.org/10.1038/nphys1274
  http://www.nature.com/nphys/journal/v5/n6/suppinfo/nphys1274\_S1.html}
  {\bibfield  {journal} {\bibinfo  {journal} {Nat. Phys.}\ }\textbf {\bibinfo
  {volume} {5}},\ \bibinfo {pages} {398} (\bibinfo {year} {2009})}\BibitemShut
  {NoStop}%
\bibitem [{\citenamefont {Henk}\ \emph {et~al.}(2012)\citenamefont {Henk},
  \citenamefont {Flieger}, \citenamefont {Maznichenko}, \citenamefont {Mertig},
  \citenamefont {Ernst}, \citenamefont {Eremeev},\ and\ \citenamefont
  {Chulkov}}]{PhysRevLett.109.076801}%
  \BibitemOpen
  \bibfield  {author} {\bibinfo {author} {\bibfnamefont {J.}~\bibnamefont
  {Henk}}, \bibinfo {author} {\bibfnamefont {M.}~\bibnamefont {Flieger}},
  \bibinfo {author} {\bibfnamefont {I.~V.}\ \bibnamefont {Maznichenko}},
  \bibinfo {author} {\bibfnamefont {I.}~\bibnamefont {Mertig}}, \bibinfo
  {author} {\bibfnamefont {A.}~\bibnamefont {Ernst}}, \bibinfo {author}
  {\bibfnamefont {S.~V.}\ \bibnamefont {Eremeev}}, \ and\ \bibinfo {author}
  {\bibfnamefont {E.~V.}\ \bibnamefont {Chulkov}},\ }\href {\doibase
  10.1103/PhysRevLett.109.076801} {\bibfield  {journal} {\bibinfo  {journal}
  {Phys. Rev. Lett.}\ }\textbf {\bibinfo {volume} {109}},\ \bibinfo {pages}
  {076801} (\bibinfo {year} {2012})}\BibitemShut {NoStop}%
\bibitem [{\citenamefont {Narayan}\ \emph {et~al.}(2014)\citenamefont
  {Narayan}, \citenamefont {Rungger}, \citenamefont {Droghetti},\ and\
  \citenamefont {Sanvito}}]{PhysRevB.90.205431}%
  \BibitemOpen
  \bibfield  {author} {\bibinfo {author} {\bibfnamefont {A.}~\bibnamefont
  {Narayan}}, \bibinfo {author} {\bibfnamefont {I.}~\bibnamefont {Rungger}},
  \bibinfo {author} {\bibfnamefont {A.}~\bibnamefont {Droghetti}}, \ and\
  \bibinfo {author} {\bibfnamefont {S.}~\bibnamefont {Sanvito}},\ }\href
  {\doibase 10.1103/PhysRevB.90.205431} {\bibfield  {journal} {\bibinfo
  {journal} {Phys. Rev. B}\ }\textbf {\bibinfo {volume} {90}},\ \bibinfo
  {pages} {205431} (\bibinfo {year} {2014})}\BibitemShut {NoStop}%
\bibitem [{\citenamefont {Narayan}\ \emph {et~al.}(2015)\citenamefont
  {Narayan}, \citenamefont {Rungger},\ and\ \citenamefont
  {Sanvito}}]{Awadhesh2015}%
  \BibitemOpen
  \bibfield  {author} {\bibinfo {author} {\bibfnamefont {A.}~\bibnamefont
  {Narayan}}, \bibinfo {author} {\bibfnamefont {I.}~\bibnamefont {Rungger}}, \
  and\ \bibinfo {author} {\bibfnamefont {S.}~\bibnamefont {Sanvito}},\ }\href
  {http://stacks.iop.org/1367-2630/17/i=3/a=033021} {\bibfield  {journal}
  {\bibinfo  {journal} {New J. Phys.}\ }\textbf {\bibinfo {volume} {17}},\
  \bibinfo {pages} {033021} (\bibinfo {year} {2015})}\BibitemShut {NoStop}%
\bibitem [{\citenamefont {Kim}\ \emph {et~al.}(2011)\citenamefont {Kim},
  \citenamefont {Ye}, \citenamefont {Kuroda}, \citenamefont {Yamada},
  \citenamefont {Krasovskii}, \citenamefont {Chulkov}, \citenamefont
  {Miyamoto}, \citenamefont {Nakatake}, \citenamefont {Okuda}, \citenamefont
  {Ueda}, \citenamefont {Shimada}, \citenamefont {Namatame}, \citenamefont
  {Taniguchi},\ and\ \citenamefont {Kimura}}]{PhysRevLett.107.056803}%
  \BibitemOpen
  \bibfield  {author} {\bibinfo {author} {\bibfnamefont {S.}~\bibnamefont
  {Kim}}, \bibinfo {author} {\bibfnamefont {M.}~\bibnamefont {Ye}}, \bibinfo
  {author} {\bibfnamefont {K.}~\bibnamefont {Kuroda}}, \bibinfo {author}
  {\bibfnamefont {Y.}~\bibnamefont {Yamada}}, \bibinfo {author} {\bibfnamefont
  {E.~E.}\ \bibnamefont {Krasovskii}}, \bibinfo {author} {\bibfnamefont
  {E.~V.}\ \bibnamefont {Chulkov}}, \bibinfo {author} {\bibfnamefont
  {K.}~\bibnamefont {Miyamoto}}, \bibinfo {author} {\bibfnamefont
  {M.}~\bibnamefont {Nakatake}}, \bibinfo {author} {\bibfnamefont
  {T.}~\bibnamefont {Okuda}}, \bibinfo {author} {\bibfnamefont
  {Y.}~\bibnamefont {Ueda}}, \bibinfo {author} {\bibfnamefont {K.}~\bibnamefont
  {Shimada}}, \bibinfo {author} {\bibfnamefont {H.}~\bibnamefont {Namatame}},
  \bibinfo {author} {\bibfnamefont {M.}~\bibnamefont {Taniguchi}}, \ and\
  \bibinfo {author} {\bibfnamefont {A.}~\bibnamefont {Kimura}},\ }\href
  {\doibase 10.1103/PhysRevLett.107.056803} {\bibfield  {journal} {\bibinfo
  {journal} {Phys. Rev. Lett.}\ }\textbf {\bibinfo {volume} {107}},\ \bibinfo
  {pages} {056803} (\bibinfo {year} {2011})}\BibitemShut {NoStop}%
\bibitem [{\citenamefont {Brahlek}\ \emph {et~al.}(2015)\citenamefont
  {Brahlek}, \citenamefont {Koirala}, \citenamefont {Bansal},\ and\
  \citenamefont {Oh}}]{Brahlek201554}%
  \BibitemOpen
  \bibfield  {author} {\bibinfo {author} {\bibfnamefont {M.}~\bibnamefont
  {Brahlek}}, \bibinfo {author} {\bibfnamefont {N.}~\bibnamefont {Koirala}},
  \bibinfo {author} {\bibfnamefont {N.}~\bibnamefont {Bansal}}, \ and\ \bibinfo
  {author} {\bibfnamefont {S.}~\bibnamefont {Oh}},\ }\href {\doibase
  http://dx.doi.org/10.1016/j.ssc.2014.10.021} {\bibfield  {journal} {\bibinfo
  {journal} {Solid State Commun.}\ }\textbf {\bibinfo {volume} {215-€"216}},\
  \bibinfo {pages} {54 } (\bibinfo {year} {2015})}\BibitemShut {NoStop}%
\bibitem [{\citenamefont {de~Vries}\ \emph {et~al.}(2017)\citenamefont
  {de~Vries}, \citenamefont {Pezzini}, \citenamefont {Meijer}, \citenamefont
  {Koirala}, \citenamefont {Salehi}, \citenamefont {Moon}, \citenamefont {Oh},
  \citenamefont {Wiedmann},\ and\ \citenamefont {Banerjee}}]{deVries2017}%
  \BibitemOpen
  \bibfield  {author} {\bibinfo {author} {\bibfnamefont {E.~K.}\ \bibnamefont
  {de~Vries}}, \bibinfo {author} {\bibfnamefont {S.}~\bibnamefont {Pezzini}},
  \bibinfo {author} {\bibfnamefont {M.~J.}\ \bibnamefont {Meijer}}, \bibinfo
  {author} {\bibfnamefont {N.}~\bibnamefont {Koirala}}, \bibinfo {author}
  {\bibfnamefont {M.}~\bibnamefont {Salehi}}, \bibinfo {author} {\bibfnamefont
  {J.}~\bibnamefont {Moon}}, \bibinfo {author} {\bibfnamefont {S.}~\bibnamefont
  {Oh}}, \bibinfo {author} {\bibfnamefont {S.}~\bibnamefont {Wiedmann}}, \ and\
  \bibinfo {author} {\bibfnamefont {T.}~\bibnamefont {Banerjee}},\ }\href
  {\doibase 10.1103/PhysRevB.96.045433} {\bibfield  {journal} {\bibinfo
  {journal} {Phys. Rev. B}\ }\textbf {\bibinfo {volume} {96}},\ \bibinfo
  {pages} {045433} (\bibinfo {year} {2017})}\BibitemShut {NoStop}%
\bibitem [{\citenamefont {Yazyev}\ \emph {et~al.}(2012)\citenamefont {Yazyev},
  \citenamefont {Kioupakis}, \citenamefont {Moore},\ and\ \citenamefont
  {Louie}}]{Louie2012}%
  \BibitemOpen
  \bibfield  {author} {\bibinfo {author} {\bibfnamefont {O.~V.}\ \bibnamefont
  {Yazyev}}, \bibinfo {author} {\bibfnamefont {E.}~\bibnamefont {Kioupakis}},
  \bibinfo {author} {\bibfnamefont {J.~E.}\ \bibnamefont {Moore}}, \ and\
  \bibinfo {author} {\bibfnamefont {S.~G.}\ \bibnamefont {Louie}},\ }\href
  {\doibase 10.1103/PhysRevB.85.161101} {\bibfield  {journal} {\bibinfo
  {journal} {Phys. Rev. B}\ }\textbf {\bibinfo {volume} {85}},\ \bibinfo
  {pages} {161101} (\bibinfo {year} {2012})}\BibitemShut {NoStop}%
\bibitem [{\citenamefont {F\"orster}\ \emph {et~al.}(2015)\citenamefont
  {F\"orster}, \citenamefont {Kr\"uger},\ and\ \citenamefont
  {Rohlfing}}]{PhysRevB.92.201404}%
  \BibitemOpen
  \bibfield  {author} {\bibinfo {author} {\bibfnamefont {T.}~\bibnamefont
  {F\"orster}}, \bibinfo {author} {\bibfnamefont {P.}~\bibnamefont {Kr\"uger}},
  \ and\ \bibinfo {author} {\bibfnamefont {M.}~\bibnamefont {Rohlfing}},\
  }\href {\doibase 10.1103/PhysRevB.92.201404} {\bibfield  {journal} {\bibinfo
  {journal} {Phys. Rev. B}\ }\textbf {\bibinfo {volume} {92}},\ \bibinfo
  {pages} {201404} (\bibinfo {year} {2015})}\BibitemShut {NoStop}%
\bibitem [{\citenamefont {F\"orster}\ \emph {et~al.}(2016)\citenamefont
  {F\"orster}, \citenamefont {Kr\"uger},\ and\ \citenamefont
  {Rohlfing}}]{PhysRevB.93.205442}%
  \BibitemOpen
  \bibfield  {author} {\bibinfo {author} {\bibfnamefont {T.}~\bibnamefont
  {F\"orster}}, \bibinfo {author} {\bibfnamefont {P.}~\bibnamefont {Kr\"uger}},
  \ and\ \bibinfo {author} {\bibfnamefont {M.}~\bibnamefont {Rohlfing}},\
  }\href {\doibase 10.1103/PhysRevB.93.205442} {\bibfield  {journal} {\bibinfo
  {journal} {Phys. Rev. B}\ }\textbf {\bibinfo {volume} {93}},\ \bibinfo
  {pages} {205442} (\bibinfo {year} {2016})}\BibitemShut {NoStop}%
\bibitem [{\citenamefont {Nechaev}\ \emph {et~al.}(2013)\citenamefont
  {Nechaev}, \citenamefont {Hatch}, \citenamefont {Bianchi}, \citenamefont
  {Guan}, \citenamefont {Friedrich}, \citenamefont {Aguilera}, \citenamefont
  {Mi}, \citenamefont {Iversen}, \citenamefont {Bl\"{u}gel}, \citenamefont
  {Hofmann},\ and\ \citenamefont {Chulkov}}]{Nechaev2013a}%
  \BibitemOpen
  \bibfield  {author} {\bibinfo {author} {\bibfnamefont {I.~A.}\ \bibnamefont
  {Nechaev}}, \bibinfo {author} {\bibfnamefont {R.~C.}\ \bibnamefont {Hatch}},
  \bibinfo {author} {\bibfnamefont {M.}~\bibnamefont {Bianchi}}, \bibinfo
  {author} {\bibfnamefont {D.}~\bibnamefont {Guan}}, \bibinfo {author}
  {\bibfnamefont {C.}~\bibnamefont {Friedrich}}, \bibinfo {author}
  {\bibfnamefont {I.}~\bibnamefont {Aguilera}}, \bibinfo {author}
  {\bibfnamefont {J.~L.}\ \bibnamefont {Mi}}, \bibinfo {author} {\bibfnamefont
  {B.~B.}\ \bibnamefont {Iversen}}, \bibinfo {author} {\bibfnamefont
  {S.}~\bibnamefont {Bl\"{u}gel}}, \bibinfo {author} {\bibfnamefont
  {P.}~\bibnamefont {Hofmann}}, \ and\ \bibinfo {author} {\bibfnamefont
  {E.~V.}\ \bibnamefont {Chulkov}},\ }\href {\doibase
  10.1103/PhysRevB.87.121111} {\bibfield  {journal} {\bibinfo  {journal} {Phys.
  Rev. B}\ }\textbf {\bibinfo {volume} {87}},\ \bibinfo {pages} {121111}
  (\bibinfo {year} {2013})}\BibitemShut {NoStop}%
\bibitem [{\citenamefont {Aguilera}\ \emph {et~al.}(2013)\citenamefont
  {Aguilera}, \citenamefont {Friedrich}, \citenamefont {Bihlmayer},\ and\
  \citenamefont {Bl\"{u}gel}}]{Aguilera2013a}%
  \BibitemOpen
  \bibfield  {author} {\bibinfo {author} {\bibfnamefont {I.}~\bibnamefont
  {Aguilera}}, \bibinfo {author} {\bibfnamefont {C.}~\bibnamefont {Friedrich}},
  \bibinfo {author} {\bibfnamefont {G.}~\bibnamefont {Bihlmayer}}, \ and\
  \bibinfo {author} {\bibfnamefont {S.}~\bibnamefont {Bl\"{u}gel}},\ }\href
  {\doibase 10.1103/PhysRevB.88.045206} {\bibfield  {journal} {\bibinfo
  {journal} {Phys. Rev. B}\ }\textbf {\bibinfo {volume} {88}},\ \bibinfo
  {pages} {045206} (\bibinfo {year} {2013})}\BibitemShut {NoStop}%
\bibitem [{\citenamefont {Mao}\ \emph {et~al.}(2011)\citenamefont {Mao},
  \citenamefont {Yamakage},\ and\ \citenamefont
  {Kuramoto}}]{PhysRevB.84.115413}%
  \BibitemOpen
  \bibfield  {author} {\bibinfo {author} {\bibfnamefont {S.}~\bibnamefont
  {Mao}}, \bibinfo {author} {\bibfnamefont {A.}~\bibnamefont {Yamakage}}, \
  and\ \bibinfo {author} {\bibfnamefont {Y.}~\bibnamefont {Kuramoto}},\ }\href
  {\doibase 10.1103/PhysRevB.84.115413} {\bibfield  {journal} {\bibinfo
  {journal} {Phys. Rev. B}\ }\textbf {\bibinfo {volume} {84}},\ \bibinfo
  {pages} {115413} (\bibinfo {year} {2011})}\BibitemShut {NoStop}%
\bibitem [{\citenamefont {Seixas}\ \emph {et~al.}(2013)\citenamefont {Seixas},
  \citenamefont {Abdalla}, \citenamefont {Schmidt}, \citenamefont {Fazzio},\
  and\ \citenamefont {Miwa}}]{Seixas2013}%
  \BibitemOpen
  \bibfield  {author} {\bibinfo {author} {\bibfnamefont {L.}~\bibnamefont
  {Seixas}}, \bibinfo {author} {\bibfnamefont {L.~B.}\ \bibnamefont {Abdalla}},
  \bibinfo {author} {\bibfnamefont {T.~M.}\ \bibnamefont {Schmidt}}, \bibinfo
  {author} {\bibfnamefont {A.}~\bibnamefont {Fazzio}}, \ and\ \bibinfo {author}
  {\bibfnamefont {R.~H.}\ \bibnamefont {Miwa}},\ }\href {\doibase
  http://dx.doi.org/10.1063/1.4773325} {\bibfield  {journal} {\bibinfo
  {journal} {J. Appl. Phys.}\ }\textbf {\bibinfo {volume} {113}},\ \bibinfo
  {eid} {023705} (\bibinfo {year} {2013})}\BibitemShut {NoStop}%
\bibitem [{\citenamefont {Yazyev}\ \emph {et~al.}(2010)\citenamefont {Yazyev},
  \citenamefont {Moore},\ and\ \citenamefont {Louie}}]{PhysRevLett.105.266806}%
  \BibitemOpen
  \bibfield  {author} {\bibinfo {author} {\bibfnamefont {O.~V.}\ \bibnamefont
  {Yazyev}}, \bibinfo {author} {\bibfnamefont {J.~E.}\ \bibnamefont {Moore}}, \
  and\ \bibinfo {author} {\bibfnamefont {S.~G.}\ \bibnamefont {Louie}},\ }\href
  {\doibase 10.1103/PhysRevLett.105.266806} {\bibfield  {journal} {\bibinfo
  {journal} {Phys. Rev. Lett.}\ }\textbf {\bibinfo {volume} {105}},\ \bibinfo
  {pages} {266806} (\bibinfo {year} {2010})}\BibitemShut {NoStop}%
\bibitem [{\citenamefont {Capelle}(2006)}]{Capelle2006bird}%
  \BibitemOpen
  \bibfield  {author} {\bibinfo {author} {\bibfnamefont {K.}~\bibnamefont
  {Capelle}},\ }\href {\doibase 10.1590/S0103-97332006000700035} {\bibfield
  {journal} {\bibinfo  {journal} {Braz. J. Phys.}\ }\textbf {\bibinfo {volume}
  {36}},\ \bibinfo {pages} {1318} (\bibinfo {year} {2006})}\BibitemShut
  {NoStop}%
\bibitem [{\citenamefont {Soler}\ \emph {et~al.}(2002)\citenamefont {Soler},
  \citenamefont {Artacho}, \citenamefont {Gale}, \citenamefont {Garc{\'\i}a},
  \citenamefont {Junquera}, \citenamefont {Ordej{\'o}n},\ and\ \citenamefont
  {S{\'a}nchez-Portal}}]{soler2002siesta}%
  \BibitemOpen
  \bibfield  {author} {\bibinfo {author} {\bibfnamefont {J.~M.}\ \bibnamefont
  {Soler}}, \bibinfo {author} {\bibfnamefont {E.}~\bibnamefont {Artacho}},
  \bibinfo {author} {\bibfnamefont {J.~D.}\ \bibnamefont {Gale}}, \bibinfo
  {author} {\bibfnamefont {A.}~\bibnamefont {Garc{\'\i}a}}, \bibinfo {author}
  {\bibfnamefont {J.}~\bibnamefont {Junquera}}, \bibinfo {author}
  {\bibfnamefont {P.}~\bibnamefont {Ordej{\'o}n}}, \ and\ \bibinfo {author}
  {\bibfnamefont {D.}~\bibnamefont {S{\'a}nchez-Portal}},\ }\href
  {http://stacks.iop.org/0953-8984/14/i=11/a=302} {\bibfield  {journal}
  {\bibinfo  {journal} {J. Phys.: Condens. Matter}\ }\textbf {\bibinfo {volume}
  {14}},\ \bibinfo {pages} {2745} (\bibinfo {year} {2002})}\BibitemShut
  {NoStop}%
\bibitem [{\citenamefont {Fern{\'a}ndez-Seivane}\ \emph
  {et~al.}(2006)\citenamefont {Fern{\'a}ndez-Seivane}, \citenamefont
  {Oliveira}, \citenamefont {Sanvito},\ and\ \citenamefont
  {Ferrer}}]{fernandez2006site}%
  \BibitemOpen
  \bibfield  {author} {\bibinfo {author} {\bibfnamefont {L.}~\bibnamefont
  {Fern{\'a}ndez-Seivane}}, \bibinfo {author} {\bibfnamefont {M.~A.}\
  \bibnamefont {Oliveira}}, \bibinfo {author} {\bibfnamefont {S.}~\bibnamefont
  {Sanvito}}, \ and\ \bibinfo {author} {\bibfnamefont {J.}~\bibnamefont
  {Ferrer}},\ }\href {\doibase 10.1088/0953-8984/19/48/489001} {\bibfield
  {journal} {\bibinfo  {journal} {J. Phys.: Condens. Matter}\ }\textbf
  {\bibinfo {volume} {18}},\ \bibinfo {pages} {7999} (\bibinfo {year}
  {2006})}\BibitemShut {NoStop}%
\bibitem [{\citenamefont {Acosta}\ \emph {et~al.}(2014)\citenamefont {Acosta},
  \citenamefont {Lima}, \citenamefont {Miwa}, \citenamefont {da~Silva},\ and\
  \citenamefont {Fazzio}}]{PhysRevB.89.155438}%
  \BibitemOpen
  \bibfield  {author} {\bibinfo {author} {\bibfnamefont {C.~M.}\ \bibnamefont
  {Acosta}}, \bibinfo {author} {\bibfnamefont {M.~P.}\ \bibnamefont {Lima}},
  \bibinfo {author} {\bibfnamefont {R.~H.}\ \bibnamefont {Miwa}}, \bibinfo
  {author} {\bibfnamefont {A.~J.~R.}\ \bibnamefont {da~Silva}}, \ and\ \bibinfo
  {author} {\bibfnamefont {A.}~\bibnamefont {Fazzio}},\ }\href {\doibase
  10.1103/PhysRevB.89.155438} {\bibfield  {journal} {\bibinfo  {journal} {Phys.
  Rev. B}\ }\textbf {\bibinfo {volume} {89}},\ \bibinfo {pages} {155438}
  (\bibinfo {year} {2014})}\BibitemShut {NoStop}%
\bibitem [{\citenamefont {Perdew}\ and\ \citenamefont
  {Zunger}(1981)}]{perdew1981self}%
  \BibitemOpen
  \bibfield  {author} {\bibinfo {author} {\bibfnamefont {J.~P.}\ \bibnamefont
  {Perdew}}\ and\ \bibinfo {author} {\bibfnamefont {A.}~\bibnamefont
  {Zunger}},\ }\href {\doibase 10.1103/PhysRevB.23.5048} {\bibfield  {journal}
  {\bibinfo  {journal} {Phys. Rev. B}\ }\textbf {\bibinfo {volume} {23}},\
  \bibinfo {pages} {5048} (\bibinfo {year} {1981})}\BibitemShut {NoStop}%
\bibitem [{Note1()}]{Note1}%
  \BibitemOpen
  \bibinfo {note} {We note that Ref.~\protect \rev@citealpnum
  {PhysRevB.82.045122} presents an $8\times 8$ Hamiltonian slightly different
  from ours, but does not study the consequences of the additional bands. The
  focus of this seminal paper is the study of $\protect \mathcal
  {H}_{1/2}({\protect \bm {k}})$.}\BibitemShut {Stop}%
\bibitem [{\citenamefont {Ebihara}\ \emph {et~al.}(2012)\citenamefont
  {Ebihara}, \citenamefont {Yada}, \citenamefont {Yamakage},\ and\
  \citenamefont {Tanaka}}]{Ebihara2012885}%
  \BibitemOpen
  \bibfield  {author} {\bibinfo {author} {\bibfnamefont {K.}~\bibnamefont
  {Ebihara}}, \bibinfo {author} {\bibfnamefont {K.}~\bibnamefont {Yada}},
  \bibinfo {author} {\bibfnamefont {A.}~\bibnamefont {Yamakage}}, \ and\
  \bibinfo {author} {\bibfnamefont {Y.}~\bibnamefont {Tanaka}},\ }\href
  {\doibase http://dx.doi.org/10.1016/j.physe.2011.12.008} {\bibfield
  {journal} {\bibinfo  {journal} {Physica E}\ }\textbf {\bibinfo {volume}
  {44}},\ \bibinfo {pages} {885 } (\bibinfo {year} {2012})}\BibitemShut
  {NoStop}%
\bibitem [{\citenamefont {Zhang}\ \emph {et~al.}(2011)\citenamefont {Zhang},
  \citenamefont {Chang}, \citenamefont {Zhang}, \citenamefont {Wen},
  \citenamefont {Feng}, \citenamefont {Li}, \citenamefont {Liu}, \citenamefont
  {He}, \citenamefont {Wang}, \citenamefont {Chen}, \citenamefont {Xue},
  \citenamefont {Ma},\ and\ \citenamefont {Wang}}]{ZhangJinsong2011}%
  \BibitemOpen
  \bibfield  {author} {\bibinfo {author} {\bibfnamefont {J.}~\bibnamefont
  {Zhang}}, \bibinfo {author} {\bibfnamefont {C.~Z.}\ \bibnamefont {Chang}},
  \bibinfo {author} {\bibfnamefont {Z.}~\bibnamefont {Zhang}}, \bibinfo
  {author} {\bibfnamefont {J.}~\bibnamefont {Wen}}, \bibinfo {author}
  {\bibfnamefont {X.}~\bibnamefont {Feng}}, \bibinfo {author} {\bibfnamefont
  {K.}~\bibnamefont {Li}}, \bibinfo {author} {\bibfnamefont {M.}~\bibnamefont
  {Liu}}, \bibinfo {author} {\bibfnamefont {K.}~\bibnamefont {He}}, \bibinfo
  {author} {\bibfnamefont {L.}~\bibnamefont {Wang}}, \bibinfo {author}
  {\bibfnamefont {X.}~\bibnamefont {Chen}}, \bibinfo {author} {\bibfnamefont
  {Q.~K.}\ \bibnamefont {Xue}}, \bibinfo {author} {\bibfnamefont
  {X.}~\bibnamefont {Ma}}, \ and\ \bibinfo {author} {\bibfnamefont
  {Y.}~\bibnamefont {Wang}},\ }\href {\doibase 10.1038/ncomms1588} {\bibfield
  {journal} {\bibinfo  {journal} {Nat. Commun.}\ }\textbf {\bibinfo {volume}
  {2}},\ \bibinfo {pages} {574} (\bibinfo {year} {2011})}\BibitemShut {NoStop}%
\bibitem [{\citenamefont {Arakane}\ \emph {et~al.}(2012)\citenamefont
  {Arakane}, \citenamefont {Sato}, \citenamefont {Souma}, \citenamefont
  {Kosaka}, \citenamefont {Nakayama}, \citenamefont {Komatsu}, \citenamefont
  {Takahashi}, \citenamefont {Ren}, \citenamefont {Segawa},\ and\ \citenamefont
  {Ando}}]{Arakane2012}%
  \BibitemOpen
  \bibfield  {author} {\bibinfo {author} {\bibfnamefont {T.}~\bibnamefont
  {Arakane}}, \bibinfo {author} {\bibfnamefont {T.}~\bibnamefont {Sato}},
  \bibinfo {author} {\bibfnamefont {S.}~\bibnamefont {Souma}}, \bibinfo
  {author} {\bibfnamefont {K.}~\bibnamefont {Kosaka}}, \bibinfo {author}
  {\bibfnamefont {K.}~\bibnamefont {Nakayama}}, \bibinfo {author}
  {\bibfnamefont {M.}~\bibnamefont {Komatsu}}, \bibinfo {author} {\bibfnamefont
  {T.}~\bibnamefont {Takahashi}}, \bibinfo {author} {\bibfnamefont
  {Z.}~\bibnamefont {Ren}}, \bibinfo {author} {\bibfnamefont {K.}~\bibnamefont
  {Segawa}}, \ and\ \bibinfo {author} {\bibfnamefont {Y.}~\bibnamefont
  {Ando}},\ }\href {http://dx.doi.org/10.1038/ncomms1639
  http://10.1038/ncomms1639
  https://www.nature.com/articles/ncomms1639{\#}supplementary-information}
  {\bibfield  {journal} {\bibinfo  {journal} {Nature Communications}\ }\textbf
  {\bibinfo {volume} {3}},\ \bibinfo {pages} {636} (\bibinfo {year}
  {2012})}\BibitemShut {NoStop}%
\bibitem [{\citenamefont {Ren}\ \emph {et~al.}(2011)\citenamefont {Ren},
  \citenamefont {Taskin}, \citenamefont {Sasaki}, \citenamefont {Segawa},\ and\
  \citenamefont {Ando}}]{PhysRevB.84.165311}%
  \BibitemOpen
  \bibfield  {author} {\bibinfo {author} {\bibfnamefont {Z.}~\bibnamefont
  {Ren}}, \bibinfo {author} {\bibfnamefont {A.~A.}\ \bibnamefont {Taskin}},
  \bibinfo {author} {\bibfnamefont {S.}~\bibnamefont {Sasaki}}, \bibinfo
  {author} {\bibfnamefont {K.}~\bibnamefont {Segawa}}, \ and\ \bibinfo {author}
  {\bibfnamefont {Y.}~\bibnamefont {Ando}},\ }\href {\doibase
  10.1103/PhysRevB.84.165311} {\bibfield  {journal} {\bibinfo  {journal} {Phys.
  Rev. B}\ }\textbf {\bibinfo {volume} {84}},\ \bibinfo {pages} {165311}
  (\bibinfo {year} {2011})}\BibitemShut {NoStop}%
\bibitem [{\citenamefont {Abdalla}\ \emph {et~al.}(2015)\citenamefont
  {Abdalla}, \citenamefont {{Padilha Jos\'e}}, \citenamefont {Schmidt},
  \citenamefont {Miwa},\ and\ \citenamefont {Fazzio}}]{Abdalla2015}%
  \BibitemOpen
  \bibfield  {author} {\bibinfo {author} {\bibfnamefont {L.~B.}\ \bibnamefont
  {Abdalla}}, \bibinfo {author} {\bibfnamefont {E.}~\bibnamefont {{Padilha
  Jos\'e}}}, \bibinfo {author} {\bibfnamefont {T.~M.}\ \bibnamefont {Schmidt}},
  \bibinfo {author} {\bibfnamefont {R.~H.}\ \bibnamefont {Miwa}}, \ and\
  \bibinfo {author} {\bibfnamefont {A.}~\bibnamefont {Fazzio}},\ }\href
  {http://stacks.iop.org/0953-8984/27/i=25/a=255501} {\bibfield  {journal}
  {\bibinfo  {journal} {J. Phys.: Condens. Matter}\ }\textbf {\bibinfo {volume}
  {27}},\ \bibinfo {pages} {255501} (\bibinfo {year} {2015})}\BibitemShut
  {NoStop}%
\bibitem [{\citenamefont {Liu}\ \emph {et~al.}(2014)\citenamefont {Liu},
  \citenamefont {Li}, \citenamefont {Rajput}, \citenamefont {Gilks},
  \citenamefont {Lari}, \citenamefont {Galindo}, \citenamefont {Weinert},
  \citenamefont {Lazarov},\ and\ \citenamefont {Li}}]{LiuY2014}%
  \BibitemOpen
  \bibfield  {author} {\bibinfo {author} {\bibfnamefont {Y.}~\bibnamefont
  {Liu}}, \bibinfo {author} {\bibfnamefont {Y.~Y.}\ \bibnamefont {Li}},
  \bibinfo {author} {\bibfnamefont {S.}~\bibnamefont {Rajput}}, \bibinfo
  {author} {\bibfnamefont {D.}~\bibnamefont {Gilks}}, \bibinfo {author}
  {\bibfnamefont {L.}~\bibnamefont {Lari}}, \bibinfo {author} {\bibfnamefont
  {P.~L.}\ \bibnamefont {Galindo}}, \bibinfo {author} {\bibfnamefont
  {M.}~\bibnamefont {Weinert}}, \bibinfo {author} {\bibfnamefont {V.~K.}\
  \bibnamefont {Lazarov}}, \ and\ \bibinfo {author} {\bibfnamefont
  {L.}~\bibnamefont {Li}},\ }\href {\doibase
  http://dx.doi.org/10.1038/nphys2898 10.1038/nphys2898} {\bibfield  {journal}
  {\bibinfo  {journal} {Nat. Phys.}\ }\textbf {\bibinfo {volume} {10}},\
  \bibinfo {pages} {294} (\bibinfo {year} {2014})}\BibitemShut {NoStop}%
\bibitem [{\citenamefont {Park}\ \emph {et~al.}(2015)\citenamefont {Park},
  \citenamefont {Chae}, \citenamefont {Jeong}, \citenamefont {Kim},
  \citenamefont {Choi}, \citenamefont {Cho}, \citenamefont {Hwang},
  \citenamefont {Bae},\ and\ \citenamefont {Kang}}]{nanolett5b00553}%
  \BibitemOpen
  \bibfield  {author} {\bibinfo {author} {\bibfnamefont {S.~H.}\ \bibnamefont
  {Park}}, \bibinfo {author} {\bibfnamefont {J.}~\bibnamefont {Chae}}, \bibinfo
  {author} {\bibfnamefont {K.~S.}\ \bibnamefont {Jeong}}, \bibinfo {author}
  {\bibfnamefont {T.~H.}\ \bibnamefont {Kim}}, \bibinfo {author} {\bibfnamefont
  {H.}~\bibnamefont {Choi}}, \bibinfo {author} {\bibfnamefont {M.~H.}\
  \bibnamefont {Cho}}, \bibinfo {author} {\bibfnamefont {I.}~\bibnamefont
  {Hwang}}, \bibinfo {author} {\bibfnamefont {M.~H.}\ \bibnamefont {Bae}}, \
  and\ \bibinfo {author} {\bibfnamefont {C.}~\bibnamefont {Kang}},\ }\href
  {\doibase 10.1021/acs.nanolett.5b00553} {\bibfield  {journal} {\bibinfo
  {journal} {Nano Lett.}\ }\textbf {\bibinfo {volume} {15}},\ \bibinfo {pages}
  {3820} (\bibinfo {year} {2015})}\BibitemShut {NoStop}%
\end{thebibliography}%

\end{document}